# Compression and Reconnection Investigations of the MagnetoPause (CRIMP): A Mission Concept to Uncover the Impact of Mesoscale Reconnection and Plasma Outflow Processes at the Dayside Magnetopause

Jason M. H. Beedle[a,#,*], Bryan C. Cline[b,#,**], Samuel T. Badman[c], Humberto Caldelas II[d,#], Kelly A. Cantwell[e,#], Alex P. Hoffmann[f], Christian Hofmann[d,#], India Jackson[g], Tre'Shunda James[f], Miguel Martínez-Ledesma[f,h], Bruno Mattos[i], Brett A. McCuen[j,#], Sophie R. Phillips[k], Bryan Reynolds[l], Julie Rolla[m], Orlando M. Romeo[n], Frances A. Staples[o,#], Michael J. Starkey[p,#], Olga P. Verkhoglyadova[m], Andres Romero-Wolf[m], Alfred E. Nash III[m]

[a]*Space Science Center, University of New Hampshire, Durham, NH, United States of America*
[b]*University of Illinois Urbana-Champaign, Urbana, IL, United States of America*
[c]*Center for Astrophysics | Harvard & Smithsonian, Cambridge, MA, United States of America*
[d]*Massachusetts Institute of Technology, Cambridge, MA, United States of America*
[e]*Dartmouth College, Hanover, NH, United States of America*
[f]*NASA Goddard Space Flight Center, Greenbelt, MD, United States of America*
[g]*Georgia State University, Atlanta, GA, United States of America*
[h]*The Catholic University of America, Washington, DC, United States of America*
[i]*Instituto Tecnológico de Aeronáutica, São Paulo, Brazil*
[j]*The Aerospace Corporation, El Segundo, CA, United States of America*
[k]*Arizona State University, Tempe, AZ, United States of America*
[l]*Remcom Inc., State College, PA, United States of America*
[m]*NASA Jet Propulsion Laboratory, California Institute of Technology, Pasadena, CA, United States of America*
[n]*Space Sciences Laboratory, University of California, Berkeley, Berkeley, CA, United States of America*
[o]*University of California, Los Angeles, Los Angeles, CA, United States of America*
[p]*Southwest Research Institute, San Antonio, TX, United States of America*

---

**Abstract**

The Compression and Reconnection Investigations of the Magnetopause (CRIMP) mission is a hypothesis-driven, Heliophysics Medium-Class Explorer (MIDEX) Announcement of Opportunity (AO) mission concept designed to study mesoscale structures and particle outflow along Earth's magnetopause using two identical spacecraft. CRIMP is designed to uncover the impact of magnetosheath mesoscale drivers, dayside magnetopause mesoscale phenomenological processes and structures, and localized plasma outflows on magnetic reconnection and the energy transfer process in the dayside magnetosphere. CRIMP accomplishes this through uniquely phased spacecraft configurations that allow multipoint, contemporaneous measurements at the magnetopause. This enables an unparalleled look at mesoscale spatial differences along the dayside magnetopause on the scale of 1-3 Earth Radii ($R_E$). Through these measurements, CRIMP will uncover how local mass density enhancements affect global reconnection, how mesoscale structures drive magnetopause dynamics, and if the magnetopause acts as a perfectly absorbing boundary for radiation belt electrons. This allows CRIMP to determine the spatial scale size, extent, and temporal evolution of energy and mass transfer processes at the magnetopause—crucial measurements to determine how the solar wind energy input to the magnetosphere is transmitted between regions and across scales.

---

[*]First Author and Corresponding Author—Science: Email: jason.beedle@unh.edu
[**]First Author and Corresponding Author—Engineering: Email: bcline3@illinois.edu
[#]Made significant contributions to the paper's formulation and composition. See Appendix A for a full list of individual contributions.

## 1. Introduction

Standing as a shield protecting the Earth, its atmosphere, and humanity's ever-expanding space-based assets, the magnetopause represents the boundary between solar wind origin plasma and the Earth's magnetic field. This boundary is supported by a current sheet and is intrinsically dynamic—fluctuating closer (farther) from Earth depending on stronger (weaker) global solar wind driving conditions. One of these global driving conditions is the direction of the solar wind's interplanetary magnetic field (IMF). When the IMF is directed southward, or in the opposite direction to the Earth's magnetic field lines, the process of magnetic reconnection can occur along the dayside or sunward facing magnetopause current sheet, enabling the explosive transfer of energy previously stored in the magnetic field into the surrounding plasma (e.g., Refs. [1, 2]). Because of this process, the dayside magnetopause acts as the primary entry point of the solar wind's energy into the magnetosphere, supplying energy to the inner magnetosphere, magnetotail, and upper atmosphere, and enabling the Dungey Cycle and its movement of mass and energy throughout the magnetosphere systems [3, 4]. Through this cycle, energy is deposited into Earth's upper atmosphere, generating the aurora and potentially causing detrimental effects to our infrastructure, including disruptions to local and national power grids and the degradation of essential orbital infrastructure (e.g., Ref. [5]). Because of the potential impact this energy transfer may have on our modern infrastructure and daily lives, understanding how magnetic reconnection is triggered, what drivers affect its efficiency, and how the resulting energetic particles from this process escape from the magnetosphere are all crucial questions that have yet to be answered.

The National Academies of Sciences, Engineering and Medicine recently highlighted the key importance of understanding this magnetospheric energy input and transfer in their congressionally-mandated 2024 Heliophysics Decadal Survey which listed as its first Priority Science Goal (PSG-1): "How is the solar wind energy input to the magnetosphere transmitted between different regions and across different scales?" [6]. To this end, the Decadal Survey underscores the need to "Determine the spatial scale size and extent and the temporal evolution of energy and mass transfer processes at the magnetopause" in its Science Goal (SG) 1a [6]. In order to address these goals, the Decadal Survey advocates for dedicated spacecraft constellations and missions designed to resolve mesoscale energy transfer processes and dynamics.

The Compression and Reconnection Investigations of the MagnetoPause (CRIMP) mission is a hypothesis-driven concept that addresses these needs by probing mesoscale magnetopause structures and magnetospheric plasmas outflows with twin spacecraft that offer in situ, spatially separated, contemporaneous measurements targeting the Decadal's PSG-1 and SG-1a by uncovering how local mass density enhancements affect global reconnection, how mesoscale structures drive magnetopause dynamics, and if the magnetopause acts as a perfectly absorbing boundary for radiation belt electrons. Figure 1 provides a visual overview of these structures and CRIMP's orbital design while Fig. 2 shows the CRIMP mission logo.

The CRIMP mission was formulated based on the specifications of the Heliophysics Explorers Program 2019 Medium-Class Explorer (MIDEX) Announcement of Opportunity (AO),[1] assuming a Principal Investigator-managed mission cost (PIMMC) cap of $250 million USD in Fiscal Year (FY) 2019 (inflation-adjusted to $301 million FY 2024 USD). This mission concept was conceived through the 2024 Heliophysics Mission Design School via the combined input of early career heliophysics researchers, early career engineers, and NASA Jet Propulsion Laboratory's (JPL) Team-X—JPL's concurrent engineering design team for space mission concept formulation [7–9].

[1] HPMIDEX19, AO NNH19ZDA013O, `https://nspires.nasaprs.com/external/solicitations/summary!init.do?solId={CE94D9F1-858E-F22E-513F-8DF6F389B4AF}&path=`, accessed 23 Jan. 2026.

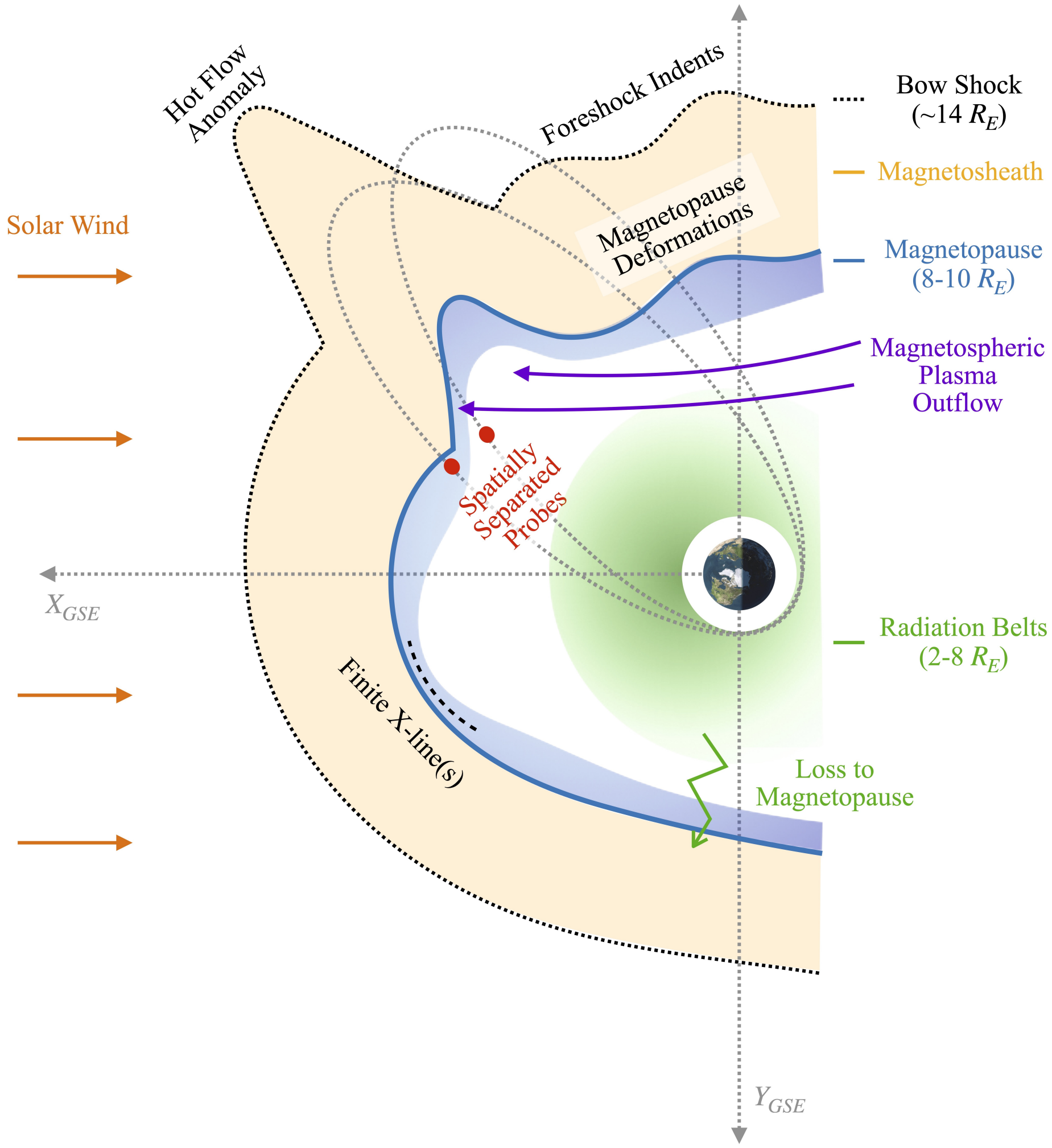

Figure 1: Figure of the dayside near-Earth environment and systems of interest for CRIMP science objectives. The Earth is represented in the center of the figure, surrounded by the radiation belts (denoted in green) which extend roughly from 2 - 8 $R_E$ (depending on solar wind driving conditions). The magnetopause boundary layer is denoted in blue and separates the shocked solar wind origin plasma in the magnetosheath (orange) from the Earth's magnetic field dominated magnetosphere. The magnetopause boundary itself is typically ∼1,000 km thick and is typically located 8-10 $R_E$ from Earth. Along this boundary, mesoscale structures, such as Kelvin-Helmholtz Vortices (i.e., large, turbulent wave-like structures), localized reconnection sites, as well as transient structures in the solar wind and along the bow shock (foreshock indents, hot flow anomalies, etc.) may cause localized magnetopause deformations along the boundary and thus impact how energy from the solar wind is able to enter into the near-Earth environment. This boundary is further complicated by outflows of plasma from the magnetosphere (denoted in purple) which can involve heavy ions from the magnetotail, radiation belt, Earth's upper atmosphere, and ring current. These particles then change the characteristics of the magnetopause plasma which can also affect this energy transfer. The local location and characteristics of the magnetopause can also determine the loss of radiation belt electrons to the solar wind. CRIMP's spatially separated, identically-phased probes are denoted in red along their twin orbital lobes denoted in grey.

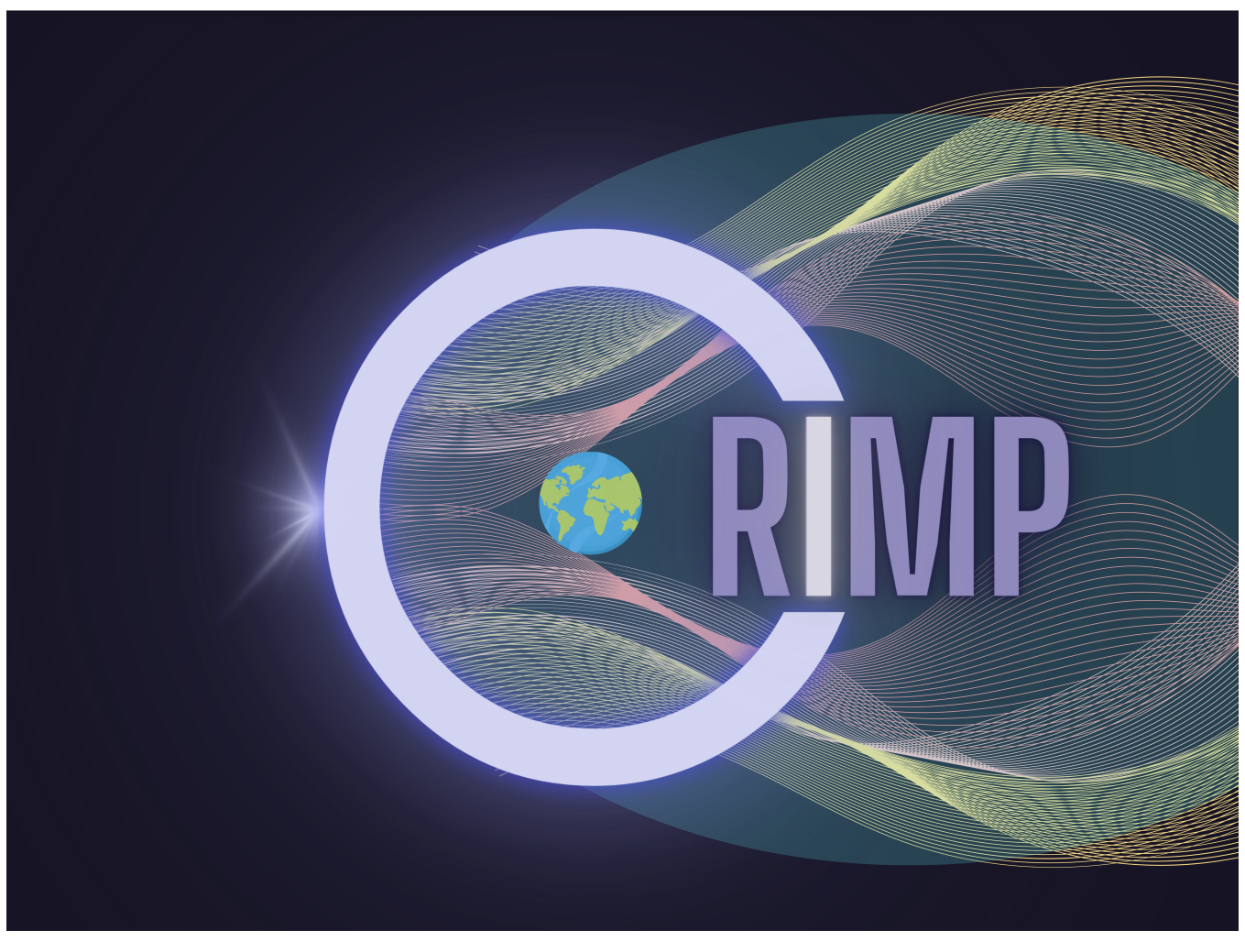

Figure 2: CRIMP Mission Logo designed by author Miguel Martínez-Ledesma.

### *1.1. Science Overview and Open Science Questions*

The magnetopause and its current system has been widely studied since it was originally proposed in the 1930s by Chapman and Ferraro [10], with the first observations of the magnetopause described in Ref. [11] from the Explorer 12 mission launched in 1961. After Explorer 12, the International Sun-Earth Explorer (ISEE) missions that launched in 1977 provided the first estimations of the magnetopause's thickness and velocity, and provided the first evidence of magnetic reconnection along the dayside magnetopause (e.g., Ref. [12]). Following ISEE, the first multi-spacecraft magnetospheric constellation mission launched with Cluster in 2000. Cluster was composed of four identical spacecraft that enabled identical in situ measurements at varying separation distances in the magnetosphere (e.g., Refs. [13, 14]). Over the 20+ years of Cluster's mission, large-scale statistics of the magnetopause's global average structure, thickness, and velocity became available with additional insights into the process of magnetic reconnection (e.g., Refs. [14, 15]). Following in Cluster's footsteps, the Time History of Events and Macroscale Interactions During Substorms (THEMIS) MIDEX Explorer class mission was launched in 2007 [16]. THEMIS is composed of five identical spacecraft that enabled a detailed investigation into magnetospheric substorm processes, as well as providing thousands of additional magnetopause crossings and refinements on magnetopause and large-scale magnetic reconnection statistical characteristics (e.g., Refs. [17–19]). Continuing the legacy of multi-spacecraft missions, NASA launched its flagship magnetospheric mission, the Magnetospheric Multiscale (MMS) mission, in 2015 to target the kinetic scale dynamics of magnetic reconnection sites and uncover their energy transfer processes [20]. MMS is composed of four identical spacecraft that travel in a precise tetrahedron formation, allowing the resolution of kinetic scale gradients in the magnetic field and plasma quantities in the magnetosphere (e.g., Ref. [21]). This, combined with MMS's unprecedented plasma measurement time resolution, allows MMS to probe the small-scale reconnection processes in the diffusion regions of magnetic reconnection and their connection with the larger scale magnetopause current structure (e.g., Refs. [21–27]). Altogether, both the global-scale structure and kinetic-scale dynamics of the magnetopause have become well established by the Cluster, THEMIS, and MMS missions. However, a critical gap remains regarding the impacts of mesoscale phenomenological structure, dynamics, and magnetospheric origin plasmas on the magnetopause and the process of magnetic reconnection [6]. This gap is exemplified by a crucial lack of mesoscale (1-5 $R_E$) measurements in the form of conjunctions between the MMS, THEMIS, and Cluster missions along the dayside magnetopause. From 2015 to 2022,[2] there are a total of 8 conjunctions (defined as $\pm 1$ $R_E$ semi-simultaneous crossings of the different mission's spacecraft at the magnetopause) between Cluster 4 and MMS1, 2 conjunctions between Cluster 4 and THEMIS A, and 19 conjunctions between MMS1 and THEMIS A at separations less than 5 $R_E$. Between 4-5 $R_E$, there are 8 conjunctions between Cluster 4 and MMS1, and 4 conjunctions between MMS1 and THEMIS A. In the range of 1-3 $R_E$, there are only $\sim$4 conjunctions in total, across all considered spacecraft pairs. Thus, even when given a considerable (seven year) time frame, and allowing for a generous margin of spacecraft proximity to the magnetopause, conjunctions at or below a mesoscale separation are rare and only occur at a rate relevant for case study applications (e.g., Ref. [28]) and not systematic explorations of these scales necessary to address decadal-level questions (e.g., PSG-1 and SG-1a) along the magnetopause. This highlights the need for a specialized mission that allows for multipoint, contemporaneous measurements at separations between 1-3 $R_E$ along the magnetopause to address this observational gap in our current and past mission coverage.

#### *1.1.1. Mesoscale Structures*

The term "mesoscale" describes scales between the small-scale (microscale) and the large-scale (macroscale). In the near-Earth environment, mesoscale is used to refer to intermediate-scale space physics that connects microphysical kinetic processes with macrophysical solar wind drivers. While the exact definition of what constitutes mesoscale is not fixed given the wide range of physical processes that it covers (see Ref. [29] and the references therein as well as Table C-2 in Ref. [6]), in this study we refer to mesoscale phenomena as involving scale sizes of approximately 1-5 $R_E$. These mesoscale phenomena form either as a result of localized solar wind conditions or interlinking processes at the bow shock, in the magnetosheath, or along the magnetopause (e.g., Refs. [29, 30]), and include (among many) solar wind Hot Flow Anomalies (HFAs), magnetopause indents, magnetosheath High Speed Jets (HSJs), Kelvin-Helmholtz instabilities (KHI), and localized X-lines/reconnection sites as depicted in Figure 1. HFAs are transient structures observed in the solar wind upstream of the bow shock that cause localized deformations of the magnetopause and are on the scale of 1 to several $R_E$ (e.g., Refs. [29, 31, 32]). Magnetopause indents and other magnetopause surface deformations are predicted to be caused by foreshock turbulence that propagates through the magnetosheath and impacts the magnetopause on scales of 1 to several $R_E$, affecting its localized structure (e.g., Refs. [30, 33]). HSJs are localized magnetosheath pressure and plasma flow enhancements that can cause $R_E$ scale magnetopause indents and potentially trigger bursty magnetic reconnection (e.g., Refs. [34–38]). KHI are velocity shear driven instabilities that develop along the flanks of the magnetopause and can form mesoscale to large-scale vortices on the magnetopause's surface that can lead to conditions favorable for magnetic reconnection (e.g., Refs. [39, 40], and references therein). Finite length X-lines (reconnection sites) can, themselves, also alter the structure of the magnetopause, potentially locally suppressing reconnection at one end of the X-line, and are also able to expand along the current sheet direction over time, dynamically affecting the magnetopause current sheet's structure (e.g., Refs. [41–43]). Altogether, these (along with many other) mesoscale phenomena are thought to fundamentally impact the magnetopause's structure and the process of magnetic reconnection along its boundary (e.g., Refs. [38, 39, 44]). However, due to the aforementioned lack of mesoscale measurements along the magnetopause, we currently rely on global simulations and rare conjunctions between existing missions to ascertain their effects, leaving many open questions regarding their nature, scale, and impact (e.g., Refs. [6, 36, 44]).

[2]Based on the CLUSTER-MMS-THEMIS Conjunctions at the magnetopause lists provided by the Cluster Science Archive. See: `https://www.cosmos.esa.int/web/csa/mms-themis-conjunctions`, accessed 23 Jan. 2026.

#### *1.1.2. Magnetospheric Ion Outflows*

While the typical mass density of magnetospheric plasma at the dayside magnetopause is low compared to the mass density of magnetosheath plasma, this is not always the case. Ion outflows from the inner magnetosphere to the dayside magnetopause can significantly increase the local mass density to rival that of the magnetosheath as exemplified in Figure 1. The main contributing ion populations include the warm plasma cloak, ring current, and plasmaspheric drainage plume (e.g., Refs. [45–52]). The warm plasma cloak consists mostly of $H^+$, $O^+$, and sometimes $He^+$ ions with energies between $\approx$10 eV to 3 keV that escape from the ionosphere at high latitudes into the magnetotail. This population convects sunward toward the dawnside magnetopause. The ring current is populated by higher energy $H^+$, $He^{++}$, $O^+$, and $He^+$ ions that have enhanced gradient drift toward the duskside magnetopause during active magnetospheric conditions. Lastly, the plasmaspheric drainage plume consists of low energy ($< 1$ eV) $H^+$, $He^+$, and sometimes $O^+$ sourced from the plasmasphere during active magnetospheric conditions. These mesoscale ion outflows can significantly enhance the local magnetospheric plasma mass density near the magnetopause and are theorized to impact the reconnection process at the magnetopause (e.g., Refs. [49, 53]). However, it is not well known how these local processes couple to global-scale energy transfer mechanisms initiated by reconnection.

#### *1.1.3. Radiation Belt Dynamics*

The most energetic population of magnetospheric plasma, the radiation belt, plays an important role in space weather. Having the ability to understand and predict the dynamics of the radiation belt is of utmost importance to protect satellite operations, communication systems, and astronaut safety. The prediction of electron loss from the radiation belt is notoriously challenging (e.g., Ref. [54]), in particular electron dropout events where nearly the entire radiation belt is lost (e.g., Ref. [55]). Magnetopause shadowing has been found to be a primary cause of radiation belt dropouts, particularly for electrons with drift paths beyond MEO altitudes (e.g., Ref. [56]). This occurs when variations in the solar wind cause the inward motion of the magnetopause, which intersects the paths of previously trapped electrons. Radiation belt research, both observational and modeling efforts, has presupposed that electrons are permanently lost through the magnetopause when their drift shell is opened to the magnetopause (see, for example, Refs. [57, 58]). However, the process by which electrons cross the magnetopause boundary is a long-standing, and yet unresolved, question in magnetospheric physics [59].

### 1.2. Mission Design Overview

CRIMP's overarching science goal is to ***understand the dynamic evolution of particle loss and the conditions that lead to magnetic reconnection along the dayside magnetopause boundary.*** To achieve this goal, CRIMP will address three primary science objectives using multipoint, contemporaneous observations spatially separated by 1-3 Earth Radii ($R_E$) along the dayside magnetopause boundary. CRIMP comprises two identical spacecraft equipped with electron, ion, and magnetic field instruments in highly elliptical (≈32.7 hour phase-synchronized) orbits. This orbital configuration will enable CRIMP to address the open science questions discussed above and improve our understanding of crucial physical processes driving the solar-wind-magnetosphere system while fitting within the constraints of a Heliophysics MIDEX AO call.

The remainder of this paper is organized as follows: Section 2 details the scientific motivation and objectives. The selected scientific instruments are described in Sec. 3. Section 4 describes the mission requirements and orbital configuration while the spacecraft subsystem design is detailed in Sec. 5. Cost analysis is provided in Sec. 6 and additional discussion is provided in Sec. 7. Finally, the paper is concluded in Sec. 8.

## 2. Scientific Motivation and Objectives

### 2.1. Objective 1: Magnetospheric Plasma Outflows and Their Effect on Magnetic Reconnection

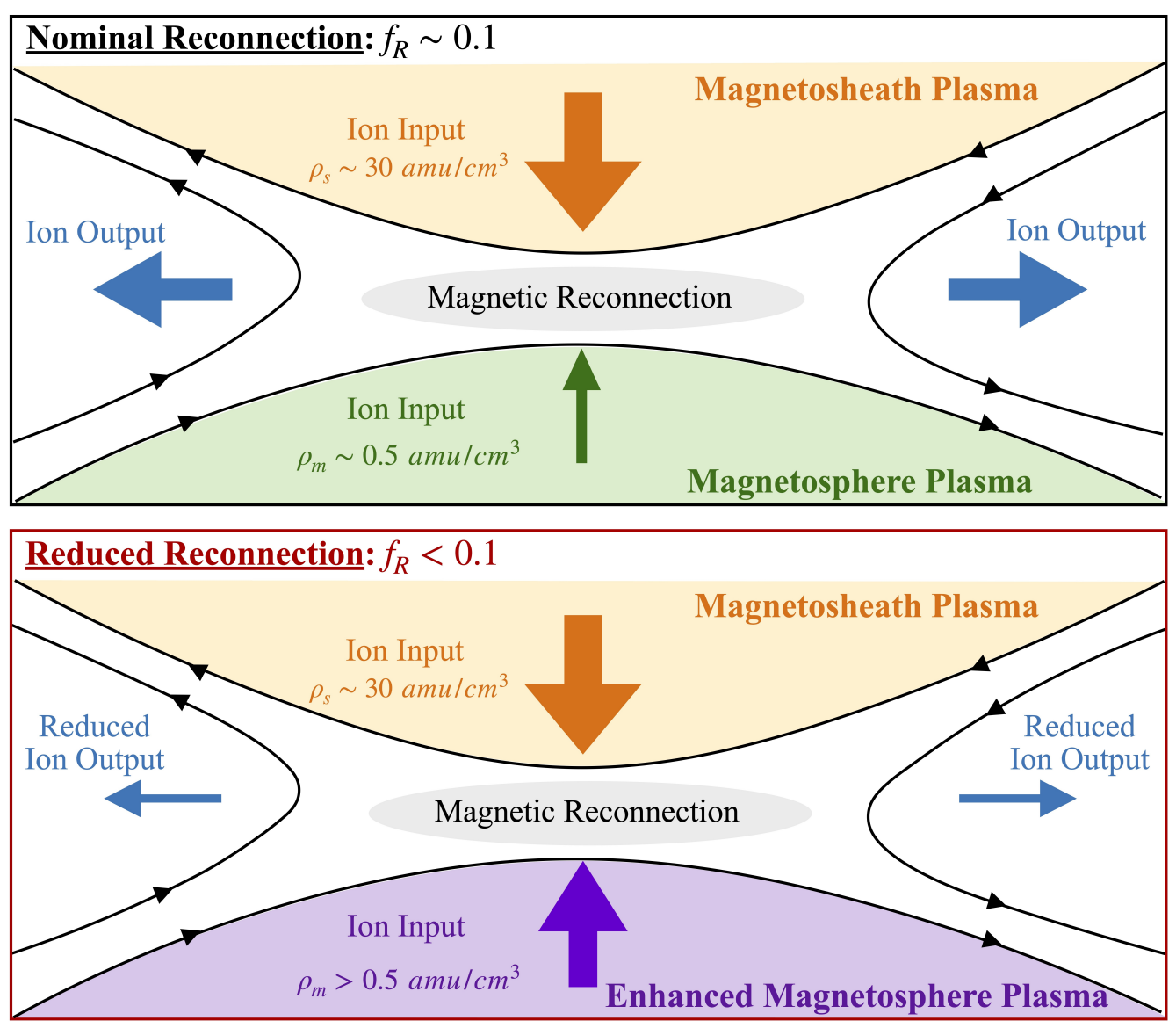

Figure 3: (Top) Schematic for typical dayside magnetic reconnection resulting in a nominal reconnection rate of $f_R \sim 0.1$. (Bottom) Schematic of dayside reconnection with a reduced reconnection rate due to the presence of heavy ion outflow from the magnetosphere, which modifies the magnetospheric plasma mass density. In the presence of significant magnetospheric ion input, the ion output from reconnection is reduced which hinders the global energy transfer from the solar wind to the magnetosphere.

The dayside reconnection rate ($f_R$, defined as the ratio of outflow to inflow ion flow speed associated with dayside reconnection) directly quantifies the rate at which energy is transferred between the magnetosheath and magnetosphere plasma populations and is key to understanding the global nature of energy flow throughout the magnetospheric system. Reconnection at the dayside magnetopause is highly asymmetric due to the differences between typical magnetospheric (with density between 0.3 - 0.5 cm$^3$ and magnetic field strength ∼50 - 60 nT) and magnetosheath plasma populations (with density between 20 - 30 cm$^3$ and magnetic field strength ∼20 - 30 nT) [60]. To account for this asymmetry (i.e., nonzero magnetospheric mass density), the authors of Ref. [61] computed a correction ($R$) to the rate derived in Refs. [60] and [62] that is given by:

$$R = (\rho_s B_m)^{1/2} / (\rho_m B_s + \rho_s B_m)^{1/2} \tag{1}$$

Here, $\rho$ is the mass density, B is the magnetic field strength, and the subscripts 's' and 'm' refer to the magnetosheath and magnetosphere, respectively. This quantity ranges from 0 to 1 and quantifies the reduction in the expected reconnection rate due to the presence of magnetospheric plasma.

In general, the global reconnection rate quantifies the overall energy input into the Earth's magnetosphere via reconnection at the dayside magnetopause. However, due to ion outflows from the inner magnetosphere (e.g., warm plasma cloak, plasmaspheric drainage plume, and ring current plasma), the local reconnection rate along the magnetopause can vary over spatial scales corresponding to these outflows (e.g., mesoscales from ∼0.5-5 $R_E$, see Refs. [49, 53]). The extent to which mesoscale ion outflows affect the reconnection rate locally and the corresponding implications for energy transfer on a global scale are not known, but remain crucial toward forming a complete picture of the interaction between the solar wind and Earth's magnetosphere. This gap motivates CRIMP's first science objective:

**Objective 1**: Determine if the global dayside reconnection rate is reduced by localized enhancements in mass density of magnetospheric plasma ($H^+$, $He^+$, $He^{++}$, $O^+$) at the magnetopause boundary.

**Physical Parameters**: To achieve this objective, CRIMP will investigate the spatial distribution of the fractional reduction of local dayside reconnection rate ($R$, defined above) along the magnetopause. By deriving this physical parameter along the magnetopause with ∼2-5 $R_E$ spatial resolution, CRIMP will determine whether dayside reconnection is unaffected by enhancements in magnetospheric mass density (i.e., $R > 95\%$)—Figure 3 top panel—or if reconnection is significantly reduced (i.e., $R < 93\%$)—Figure 3 bottom panel. These bounds follow from Ref. [48] which analyzed eight magnetopause crossings using MMS data during which enhanced magnetospheric plasma was observed. They found that for the cases of the highest magnetospheric mass density, reconnection was reduced by $> 10\%$. However, due to the low number of events, no statistical conclusions could be made which highlights the need for a dedicated mission to study this effect. In order to distinguish between the 93% and 95% thresholds, the resolution of $R$ must be $\leq 1\%$. Since $R$ is a function of magnetospheric and magnetosheath mass density and magnetic field strength, this requirement drives the measurement requirements for Science Objective 1. To account for natural plasma variability, CRIMP will compute an event-averaged value of $R$ for each magnetopause crossing and characterize its uncertainty from the observed temporal variability during the event. This uncertainty estimate will enable statistically robust discrimination between the $R > 95\%$ and $R < 93\%$ regimes.

**Observables**: To compute the correction ratio, $R$, CRIMP must measure the local ion mass density for the dominant magnetospheric and magnetosheath species $H^+$, $He^+$, $He^{++}$, and $O^+$, as well as the local magnetic field strength. An ion mass density measurement range of 0-100 amu/cm$^3$, covering the range of magnetospheric and magnetosheath populations, with $< 1.4\%$ resolution is required to achieve $< 1\%$ resolution in $R$ (determined from propagation of uncertainties assuming negligible uncertainty in magnetic field measurement). Additionally, to cover the peak energies of the various ion populations at the magnetopause, the ion mass density should be measured in an energy range between 10 eV - 30 keV (e.g., Ref. [48]). The magnetic field strength measurements should range from 1-100 nT at the magnetopause and with a resolution of $< 0.5$ nT to achieve minimal uncertainty contributions to the calculation of R. The time resolution of both the ion mass density and magnetic field measurements should be $\leq 20$ s since ion composition at the magnetopause does not change significantly on time scales $< 20$ s. These requirements drive the instrument requirements discussed in Sec. 3.

### *2.2. Objective 2: Mesoscale Phenomenological Effects on the Magnetopause*

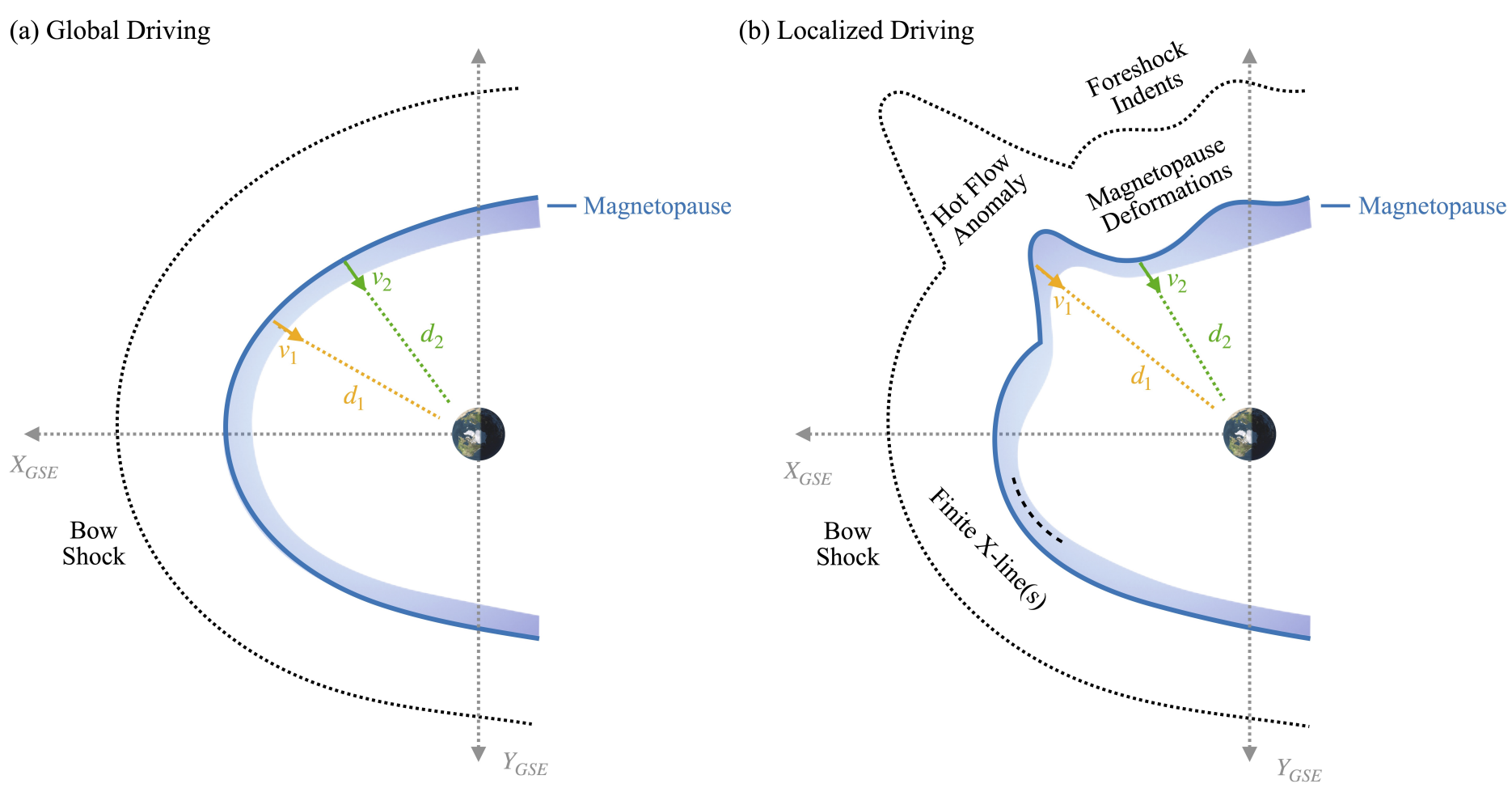

Figure 4: Diagram of two hypotheses for dayside magnetopause driving. Panel A) depicts the global driving case where the magnetopause is primarily affected by global solar wind driving conditions which cause the magnetopause to be roughly uniform (within expected statistical variation) across its surface. Thus, two mesoscale measurements made along the magnetopause's surface should return similar values for the magnetopause velocity (v) and magnetopause standoff distance from Earth (d). Panel B) depicts the localized driving case where mesoscale phenomenological processes in the solar wind, magnetosheath, and along the magnetopause lead to the development of localized magnetopause fluctuations which cause mesoscale separated measurements along the magnetopause to return significantly different values for the magnetopause velocity and standoff distance.

The magnetopause, reflecting its role as the interaction interface between the magnetosheath's shocked solar wind plasma and the Earth's magnetic field, is a highly dynamic structure. On large scales, this boundary is described by the pressure balance between the solar wind's kinetic pressure and the Earth's magnetic field pressure:

$$\rho_{sw} v_{sw}^2 \approx \frac{B_E^2}{2\mu_0}, \tag{2}$$

where $\rho_{sw}$ is the solar wind's density, $v_{sw}$ is the solar wind's velocity, $B_E$ is Earth's magnetic field strength, and $\mu_0$ is the permeability of free space (e.g., Ref. [63] and references therein). However, the local behavior of the magnetopause is also tied to embedded mesoscale transient structures in the magnetosheath and along the magnetopause boundary, as outlined in Sec. 1.1.1. Mesoscale transients (HFAs, HSJs, etc.) change the local structure of the magnetosheath plasma, causing significant deviations in the local magnetopause pressure balance. Subsequently, these modifications can cause mesoscale variations in the magnetopause boundary's structure, affecting its thickness, boundary velocity, and local distance from Earth as depicted in Figure 1 (e.g., Refs. [29, 30, 32, 38, 39]). Additionally, the formation of structures along the magnetopause itself (i.e., magnetopause indents, KHI, localized X-lines, etc.) can generate mesoscale variations in the magnetopause boundary (e.g., Refs. [43, 44, 64]). As the process of magnetic reconnection is thought to be influenced/triggered by local conditions in the magnetopause's current sheet, including by the thinning or tearing of the boundary layer [1], these mesoscale phenomena not only have a direct impact on the magnetopause, but also on the process of energy transfer throughout the magnetospheric system (e.g., Refs. [38, 39, 44]). However, to what extent these mesoscale drivers impact the magnetopause and control its dynamics, when compared with the global-scale drivers, is a relevant open question and motivates CRIMP's second science objective:

**Objective 2**: Determine if the magnetopause boundary structure is driven by global-scale solar wind interactions or by local-scale conditions in the magnetosheath.

**Physical Parameters**: Obj. 2 focuses on magnetopause structural properties as its physical parameters: i) the magnetopause velocity normal to the boundary, ii) the magnetopause thickness, and iii) the magnetopause location relative to Earth. Together, these attributes will determine whether the magnetopause is driven by a) global scale phenomenological scales, where magnetopause structural properties are constant within mesoscale phenomena (1-5 $R_E$, as shown in Figure 4a) or b) local-scales where magnetopause structural properties vary within mesoscales (Figure 4b).

**Observables**: To estimate the magnetopause structure, vector magnetic field observations in the nominal magnetospheric range of 1-100 nT with a resolution of 0.5 nT are required in order to achieve negligible uncertainty in the normal magnetopause velocity coordinate determination. Additionally, the ion particle flux has to be resolved by energy and direction in order to detect the primary ion particle populations in the energy range 100 eV - 10 kev to determine the primarily ion-gyroradius scale in the magnetopause boundary (e.g., Refs. [48, 65, 66]). These observations are required to be made with an energy resolution of $\Delta E/E < 30\%$ to resolve energy range differences between solar wind origin magnetosheath plasma and magnetospheric plasma populations. Additionally, as the typical magnetopause boundary has a thickness of $\sim$700 km and a typical speed of 50 km/s [15, 19, 26, 67–69], the ion particle fluxes have to be measured with a time resolution of 2.3 seconds in order to make at least three plasma measurements inside the nominal magnetopause. These measurement requirements help drive the instrument requirements detailed in Sec. 3.

### *2.3. Objective 3: Radiation Belt Electron Loss Dynamics*

Determining the physical mechanisms that produce outer radiation belt electron loss is one of the most pressing questions in inner magnetospheric physics. Ultra-relativistic electrons may either be lost to the atmosphere via atmospheric precipitation or to the magnetopause by magnetopause shadowing. While the process of electron loss into the upper atmosphere is fairly well defined through observations of Coulomb scattering (e.g., Ref. [70]), no measurements have yet been made for electron absorption by the magnetopause. To date, magnetopause losses have only been inferred from a depletion of trapped particles at high radial distances, coinciding with a modeled or observed magnetopause compression (e.g., Refs. [71, 72]). Therefore, this leaves the following relevant question: By what process do ultra-relativistic electrons cross the magnetopause boundary? It is possible that gradient drift forces across the tangential discontinuity in the magnetic field could transport electrons into the magnetosheath, or

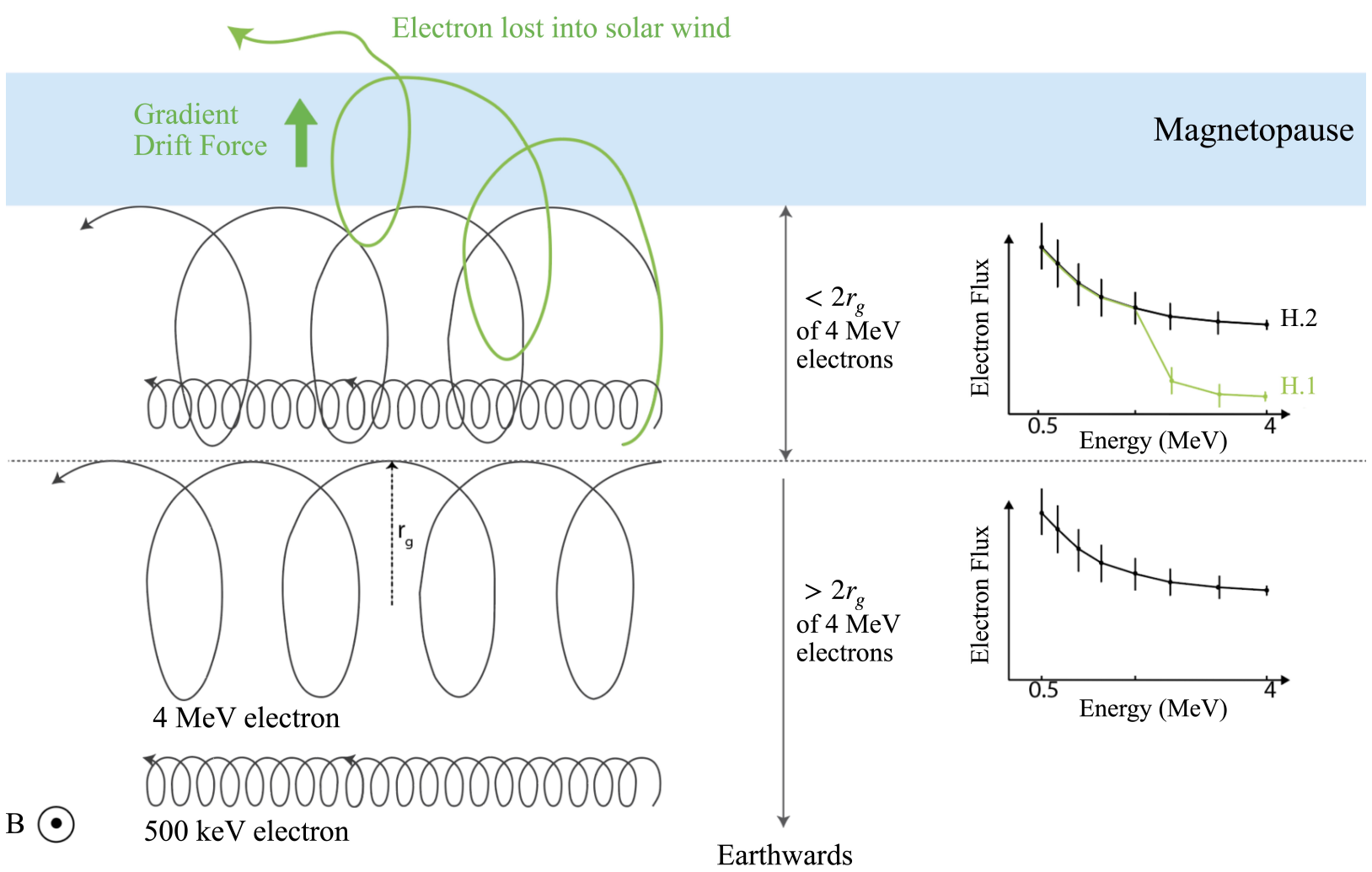

Figure 5: Diagram of ultra-relativistic electron loss to the magnetopause through gradient drift. The magnetopause is represented by the blue shaded area, with the magnetospheric magnetic field oriented perpendicular to the page shown below. The graphs on the right show the expected spectra of equatorial electron flux in the drifting region (top) and trapped region (bottom), while the diagrams on the left show the trajectories of electrons with different energies at these regions. Trapped electron motions at 500 keV and 4 MeV energies are depicted by black lines, in the electron drift region ($< 2\ r_g$ of magnetopause, above dashed line) and trapped region ($> 2\ r_g$ within magnetopause, below dashed line). Electrons lost through gradient drift are depicted in green and highlighted as H.1 in the spectra graph, and H.2 representing the null hypothesis.

electrons could stream along merged field lines into the magnetosheath following reconnection at the magnetopause (Figure 5). This intriguing question motivates the third science objective of CRIMP:

**Objective 3**: Determine if the loss of ultra-relativistic radiation belt electrons across the dayside magnetopause is due to the gradient drift of electrons across the magnetic discontinuity or magnetopause dynamics.

**Physical Parameters**: In order to study radiation belt electrons approaching the magnetopause, Obj. 3 considers electron energy spectra in the 300 keV – 4 MeV energy range at nominal radial magnetopause distances ($> 8\ R_E$) where the electron perpendicular energy ($E_\perp$) can be related to the electron gyroradius ($r_g$) and the magnetic field strength ($B$) using the following formula: $E_\perp \propto r_g \times B$. This energy range is motivated by the deceleration of ultra-relativistic radiation belt electrons (500 keV - multi-MeV at $< 4\ R_E$) which, when transported to radial distances greater than 8 $R_E$, decelerate to energies in the range of 300 keV - 4 MeV (see, for example, Refs. [56, 73]). For electrons with perpendicular energies in this 300 keV – 4 MeV range at a nominal expanded magnetopause location of 10 $R_E$, their gyroradii will be of the order of 26-95 km.

**Observables**: To detect these radiation belt electrons, it is required to measure energy flux resolved both in energy and spatial location. For the flux range, the measurement ranges between $10^1 - 10^8\ cm^{-2}s^{-1}sr^{-1}MeV^{-1}$ (or the minimum flux predicted in Ref. [74] to the maximum flux observations from the Van Allen Probes [29]). The pitch angle should be between $90^\circ \pm 20^\circ$ to measure particles with energies perpendicular to the local magnetic field in the energy range of 300 keV - 4 MeV. To resolve these particles, an energy resolution of $\Delta E/E < 60\%$ with four energy bins is required to resolve differences in the electron energy flux spectrum when approaching the magnetopause boundary. These measurements are required to be obtained at $\pm$ 2 $R_E$ on either side of the magnetopause [75] with a spatial resolution of 150 km ($\sim$2 $r_g$ for 1 MeV electrons) and a time resolution of 4 seconds to resolve the electron flux within 150 km of the magnetopause (which moves at normal velocity of $\sim$38.5 km/s, see Ref. [76]). These measurement requirements drive the instrument requirements detailed in Sec. 3.

## 3. Instruments

Each of CRIMP's spacecraft is equipped with an identical set of electron, ion, and magnetic field instruments that enable the observations necessary to address Objectives 1-3. To demonstrate the feasibility of this concept, previously flown, high technology readiness level (TRL) instruments were selected. However, the utilization of updated or purpose-built instrument designs with improved instrument sensitivities, power requirements, and fields of view is actively being explored.

### *3.1. Magnetic Field Measurements*

Common to all three of CRIMP's science objectives is the need for accurate magnetic field measurements. The strictest instrument requirements imposed on the magnetic field measurements come from Science Objective 2, which are indicated in Table 1. The digital fluxgate magnetometer (MAG) which was previously flown on THEMIS [77] exceeds all of these requirements and offers a suitable option for the CRIMP mission.

### *3.2. Particle Measurements - Electrons*

Science Objective 3 requires measurements of high-energy (300 keV - 4 MeV) electron flux with local pitch angles of $90^\circ \pm 20^\circ$ in order to determine how ultra-relativistic radiation belt electrons are lost at the magnetopause. Given the observational requirements discussed in Sec. 2, the Relativistic Electron-Proton Telescope integrated little experiment (REPTile, see Refs. [78, 79]) was identified as a suitable instrument to provide these measurements. REPTile is a small, scaled-down version of the energetic particle spectrometer Relativistic Electron-Proton Telescope (REPT; see Ref. [80]) instrument which was previously flown on the Van Allen probes. The instrument comprises a stack of solid state detectors along with a collimation aperture and penetrating radiation shielding. To maintain a field of view of the electrons that are perpendicular to the Earth's magnetic field ($90^\circ \pm 20^\circ$), REPTile must be mounted on a mechanical platform which slowly adjusts the pointing of REPTile to account for seasonal differences between the Earth's magnetic equatorial plane and the ecliptic plane.

CRIMP's energetic electron instrument requirements along with the performance of the REPTile instrument are shown in Table 1. REPTile exceeds all of the instrument requirements, thus making it a suitable option to fly on CRIMP. Since the REPTile instrument has not previously been used to measure energetic electron populations in the Earth's outer magnetosphere, a quantitative assessment of its ability to discriminate between hypotheses H.1 and H.2 should be undertaken as the design is matured. This should include forward modelling of the expected electron energy spectra under each hypothesis convolved with the REPTile instrument response, providing a quantitative demonstration of the instrument's capability to resolve the proposed science objectives.

### 3.3. Particle Measurements - Ions

Science Objectives 1 and 2 both require measurements of ion particle distributions to achieve science closure. The ion mass density is computed as the first moment of the 3D ion velocity distribution function (VDF), whereas the normal magnetopause velocity is computed using ion flux resolved by energy and direction. To achieve these measurements, CRIMP proposes the use of two separate ion instruments.

For Objective 1, since species separation is required, CRIMP plans to fly an instrument similar to the Hot Plasma Composition Analyzer (HPCA) onboard the MMS mission [81]. HPCA consists of a toroidal electrostatic analyzer in series with an optically coupled time-of-flight system, which provides measurement of an ions energy-per-charge (E/q) and mass-per-charge (m/q), respectively. The collective measurements, along with angular binning based on the spin of the MMS spacecraft, allows HPCA to measure species-resolved 3D VDFs for $H^+$, $He^+$, $He^{++}$, $O^+$, and $O^{++}$. The corresponding instrument requirements, derived from the requirements set on the observables, are shown in Table 1 along with the projected performance of the HPCA instrument. HPCA is clearly well suited for the CRIMP mission since it outperforms the instrument requirements for Objective 1.

For Objective 2, species separation is not needed. However, the time resolution requirement of 2.3 s is too fast for HPCA to achieve (which relies on the spacecraft spin period to achieve a full 3D VDF measurement; see Sec. 5.3). Thus, CRIMP proposes to fly an instrument suite composed of two ion electrostatic analyzers (ESA) similar to those flown on the THEMIS mission [82]. The ESA is a top hat back-to-back pair of hemispherical analyzers that provide measurements of ion and electron 3D VDFs. With the two ESAs mounted on opposite sides of the spinning CRIMP spacecraft, a time resolution of 1.9 s for a full 3D VDF measurement can be achieved. The instrument requirements and projected performances of the THEMIS ESA are shown in Table 1. Again, the THEMIS ESA exceeds nearly all of the CRIMP instrument requirements, which makes it an attractive candidate for the CRIMP mission.

### 3.4. Data Sufficiency and Analysis

To ensure that the CRIMP mission can achieve closure on each science objective, we estimate the mission measurement capabilities and compare this to the number of measurements required to achieve statistically significant results.

#### 3.4.1. Objective 1

Objective 1 requires contemporaneous measurements near the magnetopause with a ~2-3 $R_E$ separation accompanied by an ion outflow event from the magnetosphere. Based on Ref. [83], we assume an occurrence rate for magnetospheric ion outflows with densities $< 1$ cm$^{-3}$ to be ~30%. To cover a 30 $R_E$ extent with 3 $R_E$ resolution along the dayside magnetopause, CRIMP needs to make at least 10 measurements. Complementing this with the need to observe an ion outflow event, we estimate that the required number of observations is $10/0.3 \approx 33$ to achieve closure on Science Objective 1. To estimate the mission capability, we treat the magnetopause crossing events as a binomial probability distribution, where a success corresponds to a successful contemporaneous measurement of the magnetopause accompanied with an ion outflow and during conditions that are favorable toward magnetopause reconnection (i.e., $B_z < 0$). With the assumption that the probability of $B_z < 0$ is ≈50%, we compute the binomial probability of success to be $p = (0.5) \times (0.3) = 0.15$. Next we compute the total sample size for the binomial distribution, which is estimated as the minimum number of contemporaneous dayside magnetopause crossings over the course of the entire 2-year mission. For an orbital period of 32.7 hours (see Sec. 4.3) and only including dayside orbits (i.e., apogee is on the dayside within ± 70° of the Earth-Sun line, ≈38% of the time), we estimate the number of "good" orbits to be $N_{orb} = 2 \times 365 \times 0.38 \times (24/32.7) \approx 203$. Next, since we get at least two contemporaneous observations for each orbit (during outbound and inbound leg), the total sample size becomes $N = 2 \times N_{orb} = 406$. Note that this is a lower bound because in many cases we can expect multiple magnetopause encounters during an outbound/inbound leg of an orbit due to magnetopause motion. Lastly, for a binomial probability distribution with probability of success, $p = 0.15$, and number of events, $N = 406$, we can expect ≈50 successful events at a 95% confidence level. Since the mission capability (≈50) is significantly larger than the required number of observations (≈33) CRIMP is expected to acquire sufficient data to close on Science Objective 1.

Table 1: CRIMP Instrument Requirements and Projected Performance

| Instrument | Science Objective(s) | Instrument Requirements | Projected Performance |
|---|---|---|---|
| MAG | 1, 2, 3 | Range: 1–100 nT | 0.1–25000 nT |
| | | Resolution: 0.5 nT | Resolution: 0.25 nT |
| | | Time Resolution: 2.3 s | Time Resolution: 0.2 s |
| REPTile | 3 | Energy Range: 300 keV–4 MeV | Energy Range: 250 keV–6 MeV |
| | | Energy Resolution: $\Delta E/E < 0.6$ | Energy Resolution: $\Delta E/E < 0.07$–0.22 |
| | | Time Resolution: 4 s | Time Resolution: 1 s |
| | | FOV: 40° | FOV: 52° |
| HPCA | 1 | Mass Range: 1–16 amu/q | Mass Range: 1–28 amu/q |
| | | Mass Resolution: $M/dM > 2$ | Mass Resolution: $M/dM > 5$ |
| | | Energy Range: 10 eV–30 keV | Energy Range: 1 eV–40 keV |
| | | Energy Resolution: $\Delta E/E < 0.3$ | Energy Resolution: $\Delta E/E < 0.2$ |
| | | Time Resolution: 20 s | Time Resolution: 12 s |
| | | FOV: $4\pi$ sr | FOV: $4\pi$ sr |
| | | iFOV: $20 \times 360$° | iFOV: $11.25 \times 360$° |
| ESA | 2, 3 | Energy Range: 100 eV–10 keV | Energy Range: 7 eV–25 keV |
| | | Energy Resolution: $\Delta E/E < 0.3$ | Energy Resolution: $\Delta E/E < 0.19$ |
| | | Time Resolution: 2.3 s | Time Resolution: 1.9 s |
| | | FOV: $4\pi$ sr | FOV: $4\pi$ sr |
| | | iFOV: 180° | iFOV: 180° |

#### *3.4.2. Objective 2*

Similar to Objective 1, Objective 2 requires contemporaneous observations along the extent of the dayside magnetopause (≈30 $R_E$) with ≈2-3 $R_E$ separation. However, Objective 2 does not require the simultaneous presence of an ion outflow event. Thus, Objective 2 requires a minimum of 10 measurements to achieve science closure. Similarly, since Objective 2 does not depend on the occurrence of ion outflows or the sign of $B_z$, the mission capability is simply the minimum number of contemporaneous magnetopause encounters within ± 70 ° from the Earth-Sun line, which was computed above to be ≈406 observations. Here, the mission capability greatly exceeds the required number of observations providing high confidence that CRIMP will achieve science closure on Objective 2.

#### *3.4.3. Objective 3*

Objective 3 requires observations of the magnetopause to be made during both compressed and uncompressed magnetopause conditions. For redundancy, we require at least three observations of each group, which results in a total of six required observations. Since the uncompressed magnetopause conditions are very common, we estimate the mission capability to observe the magnetopause during compressed conditions. For this we consider compressions due to either stream interaction regions (SIRs) or interplanetary coronal mass ejections (ICMEs). The yearly occurrence rate during the descending phase of the solar cycle for SIR-induced geomagnetic storms is estimated based on Ref. [84] as $r_{SIR} = (24\, hours/SIR) \times (33\, SIR/year) = 0.09$. The occurrence rate for ICME-induced storms was estimated based on Ref. [85] as $r_{ICME} = (15\, hours/storm) \times (32\, storms/year) = 0.055$. Thus, the total occurrence rate of compressed magnetopause conditions is $r = r_{ICME} + r_{SIR}$. To estimate the mission capability for Objective 3, we model the observation occurrences as a binomial distribution with probability of success $p = r \times 0.38 = 0.055$. Here, the factor of 0.38 accounts for the time that the CRIMP orbits spend on the dayside within ± 70° from the Earth-Sun line (see Sec. 4.3). The total number of events is estimated as the total number of single-point magnetopause observations over the course of the mission, which is $N = 406 \times 2 = 812$. Thus, the expected number of successful observations with 95% confidence is ≈35. Once again, the mission capability significantly exceeds the required number of observations and so CRIMP will achieve science closure to Objective 3 with high confidence.

## 4. Mission Design

### *4.1. Overview*

The requirements of the AO and the science objectives detailed in Sec. 2 directly drove the mission, spacecraft, ground system, and operations requirements. The Mission Traceability Matrix (MTM), presented in Table 2, sum-marizes these requirements. The impacts of these requirements on the spacecraft design are discussed in Sec. 5. The requirement of cotemporal, spatially distributed measurements necessitated the use of two spacecraft, each in its own orbit. The spacecraft must be phased to cross the nominal magnetopause simultaneously in order to address the scientific objectives.

The primary mission duration was assumed to be two years. The following subsections detail the science mission profile, orbital configuration, and the launch vehicle compatibility.

Table 2: CRIMP Mission Traceability Matrix

| Mission Requirements | Mission Design Requirements | Spacecraft Requirements | Ground System Requirements | Operations Requirements |
|---|---|---|---|---|
| • Two spacecraft<br>• Orbit: Perigee $< 5$ $R_E$, Apogee 12-15 $R_E$<br>• Orbital plane: $\pm$ 20° with respect to the ecliptic plane<br>• Spacecraft separation: 2-5 $R_E$ at 8-10 $R_E$<br>• Spacecraft spin rate: $\geq$ 15 rpm<br>• Spin axis accuracy: $\pm$ 1°<br>• Spin axis knowledge: 0.1°<br>• Position knowledge: 0.1 $R_E$<br>• Data volume to downlink: 0.34 Gbits/orbit/spacecraft | • Launch vehicle type: Government furnished equipment<br>• Mission length: 2 years<br>• Orbit type: geocentric elliptical<br>• Orbit radii: Perigee $< 5$ $R_E$, Apogee 12-15 $R_E$<br>• Orbital plane: $\pm$ 20° with respect to the ecliptic plane<br>• Two separate orbits<br>• Spacecraft linear separation: 2-5 $R_E$ at radial distances of 8-10 $R_E$ along the magnetopause<br>• Phasing: 0.5 $R_E$ in arc length at a radial distance of 10 $R_E$ | • Spin stabilized: longitudinal axis perpendicular to ecliptic plane within $\pm$ 1°<br>• Pointing knowledge: 0.1°<br>• Spin rate: $\geq$ 15 rpm<br>• Mass allocation: 282.1 kg dry, 342.7 kg wet<br>• Downlink band: X<br>• Downlink data rate: 300 kbps<br>• Uplink band: X<br>• Uplink data rate: 2 kbps | • Data volume downlink per orbit: 0.34 Gbits/orbit/spacecraft<br>• Passes per orbit: 1<br>• Pass duration: 30 min<br>• Downlink band: X<br>• Downlink data rate: 300 kbits/sec<br>• Uplink band: X<br>• Uplink data rate: 2 kbits/sec<br>• MOS location: JPL<br>• SDS location: JPL | • Commanding frequency: X-band<br>• Ephemeris accuracy and update frequency: X-band<br>• Position knowledge: 0.1 $R_E$ |

### 4.2. Science Mission Profile

The scientific region of interest (ROI) is primarily between 6 $R_E$ and 11 $R_E$ (i.e., the nominal magnetopause location given by the Shue model [67]) while the spacecraft are within ±70 degrees of the Sun-Earth line on the dayside magnetopause. Data will be actively collected when the spacecraft are in this region. This ensures the generated data volume is feasible for downlinking while meeting the baseline requirements of all three scientific objectives. Downlinking occurs over approximately 30 minutes near periapsis. Additional details regarding the communications and downlinking strategy are provided in Sec. 5.4. Figure 6a schematically displays the region of interest in relation to the orbit of a single spacecraft. The green solid arc denotes the nominal magnetopause. Figure 6b displays the data generation rate and downlinking rate as a function of the orbital position. The vertical dashed lines denote the times in which the spacecraft are within the scientific ROI. Note that some telemetry data will be generated throughout the orbit. This, however, occurs at a much lower rate than during science operations. This ROI configuration allows for the measurements necessary to address Objectives 1-3 on the dayside magnetopause (±70 degrees of the Sun-Earth line) and thus achieve the baseline mission.

There is further potential to obtain data outside CRIMP's dayside-focused science objectives by adjusting the ROI to include larger radial distances for observations made on the flank magnetopause, or by parsing through lower resolution data over the orbit to identify relevant science intervals. The latter could be done using a scientist-in-the-loop (SITL) method and/or new machine learning algorithms (e.g., Ref. [86]). Doing so would enable a "guest investigator" program (see Sec. 7.2) that has the potential to address a variety of magnetotail, bow shock, and magnetosheath investigations.

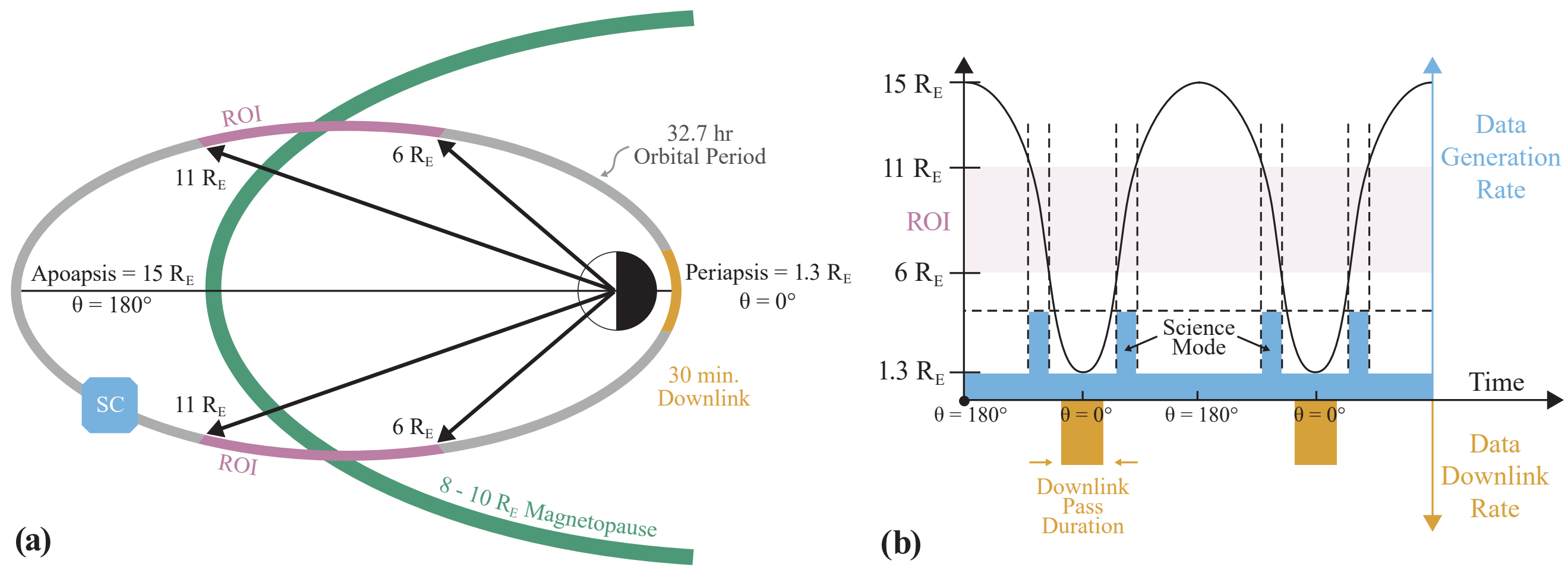

Figure 6: (a) Schematic showing CRIMP's elliptical orbit (grey line, one spacecraft) relative to the nominal magnetopause position (green line). The pink and yellow highlighted portions of the orbit indicate the portion(s) of CRIMP's orbit corresponding to the science ROI and data downlink, respectively. (b) Schematic showing CRIMP's data generation and downlink rates throughout the course of two orbits (orbital radius indicated by the black curve as a function of time).

### 4.3. Orbital Configuration

The mission requirements necessitate two spacecraft, each placed in their own orbit (or "lobe"), with identical phasing. The nominal science orbit has a semimajor axis of 51,890 km, an eccentricity of 0.84 (corresponding to an orbit between 1.3 and 15 $R_E$), and is inclined 22° relative to the Earth's equatorial plane. This trajectory has an orbital period of approximately 32.7 hr. The initial right ascension of the ascending node will be selected to approximately align the orbital plane with the ecliptic plane while the initial argument of periapsis will be selected to place the apoapsis of both orbits on the dayside. Both of these parameters are launch date-dependent. Nominally, the argument of periapsis for the two lobes are to be separated by 15°. This orbital configuration ensures adequate spacecraft separation to achieve the science objectives even when the orbits are ±70° from the Earth-Sun line (assuming a nominal magnetopause location and shape; see Ref. [67]). A two-dimensional projection of the nominal orbital configuration is shown in Figure 7.

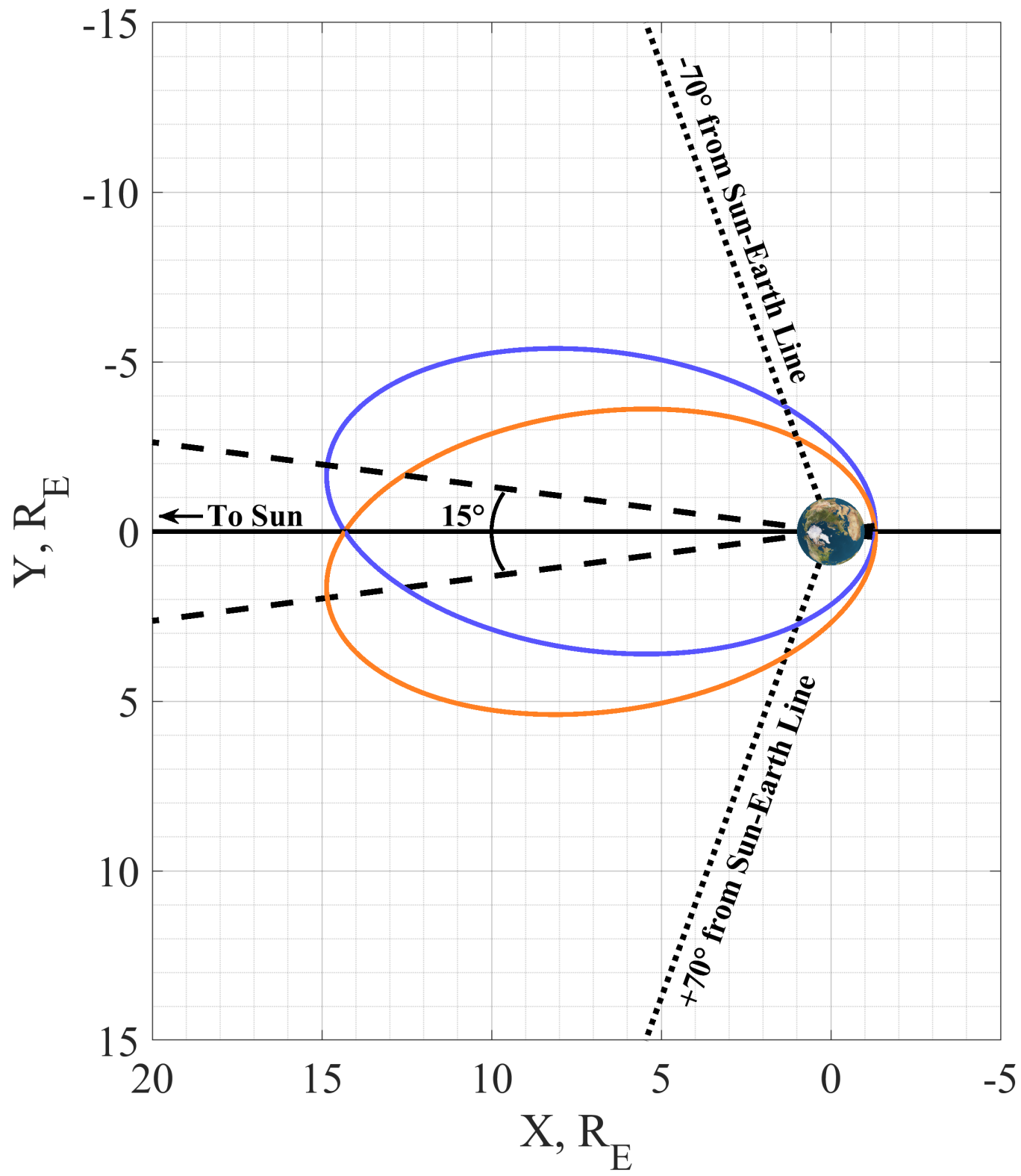

Figure 7: Two-dimensional projection of the nominal CRIMP orbital configuration in the geocentric solar ecliptic (GSE) reference frame. The orbits of the two spacecraft are represented by the blue and orange ellipses, respectively. The line of nodes for each orbit is denoted by the dashed black lines. They are separated by 15°. The black dotted lines represent ±70° from the Sun-Earth line (solid black line on the horizontal axis). The direction to the Sun is noted on the left of the figure.

Such an orbital configuration will be achieved via four total propulsive burns (two sets per spacecraft). Both spacecraft will be injected by the launch vehicle into an orbit with a periapsis of 6,569 km, apoapsis of 85,720 km, and with the semimajor axis (i.e., the line of apsides) rotated 90° relative to the Earth-Sun line on the dusk side. The first set of burns will move the argument of periapsis for both orbits to an intermediate value and raise periapsis. The second set of burns will raise the apoapsis to the desired science orbit value of 95,670 km (15 $R_E$) and increase the argument of periapsis again to the final value so that each spacecraft's argument of periapsis is separated by 15° relative to each other. These sets of burns are designed to minimize the amount of required propellant for both orbital altitude and argument of periapsis requirements for the spacecraft and fully leverage the capabilities of the launch vehicle.

The launch and early operations phase of the mission is anticipated to last one month. Three weeks will be used to perform initial system checkouts and transition to the science orbit, while one week will be dedicated to commissioning the scientific instruments. Injecting the spacecraft into the first orbit on the dusk side allows for adequate time for orbit raising and orientation burns as well as initial system checkouts.

The primary mission duration will be two years. During that time, the orbits will slowly precess with respect to the Earth due to non-spherical gravity perturbations (e.g., the Earth's $J_2$ effect) and third-body gravitation (principally from the Sun and Moon). These effects result in the line of apsides rotating on the order of ≈25° in inertial space over one year. This rotation, in addition to the Earth's motion around the Sun, results in the orbits spending part of each year outside the scientific region of interest. In total, approximately 277 days of the mission's two years (≈38%) will be spent with both orbits within ±70° of the Sun-Earth line. This is sufficient to address all three science objectives (see Sec. 3.4). The remaining portion of the primary mission may potentially be used for a guest investigator program. In this program, scientists may submit proposals to use the CRIMP instrumentation to investigate science questions outside of the scope of the CRIMP objectives.

Significant station-keeping maneuvers are not anticipated to be required because the orbits of both spacecraft evolve similarly (due to non-spherical gravity and third-body perturbations) throughout the mission. This results in the relative spacing of the spacecraft remaining nearly constant. However, 0.03 km/s of $\Delta V$ have been allocated for any required station-keeping maneuvers. Both spacecraft will lower their orbit periapsis to 50 km (requiring a $\Delta V$ of 0.09 km/s) at the end of the mission to utilize atmospheric drag for deorbiting purposes. Both station-keeping and deorbit propellant requirements are included in the total mission $\Delta V$ budget (see Sec. 5.2).

### *4.4. Launch Services and Launch Vehicle Compatibility*

The spacecraft were assumed to launch from the NASA's John F. Kennedy Space Center (Merritt Island, Florida, USA) with a required characteristic energy ($C_3$) of -9.13 km$^2$/s$^2$. For this AO, the launch vehicle is government furnished equipment and thus it is not included in the PIMMC. The spacecraft were assumed to be integrated on an Evolved Expendable Launch Vehicle (EELV) Secondary Payload Adapter (ESPA) Grande Ring within a 4.6 m diameter fairing. This leaves two remaining attachments to the ESPA Grande Ring that could potentially be used for multiple rideshare opportunities. Three renderings of the spacecraft within the fairing are shown in Figure 8. The fairing size and required $C_3$ are within the limits specified by the Launch Services Program Information Summary associated with the 2019 AO.

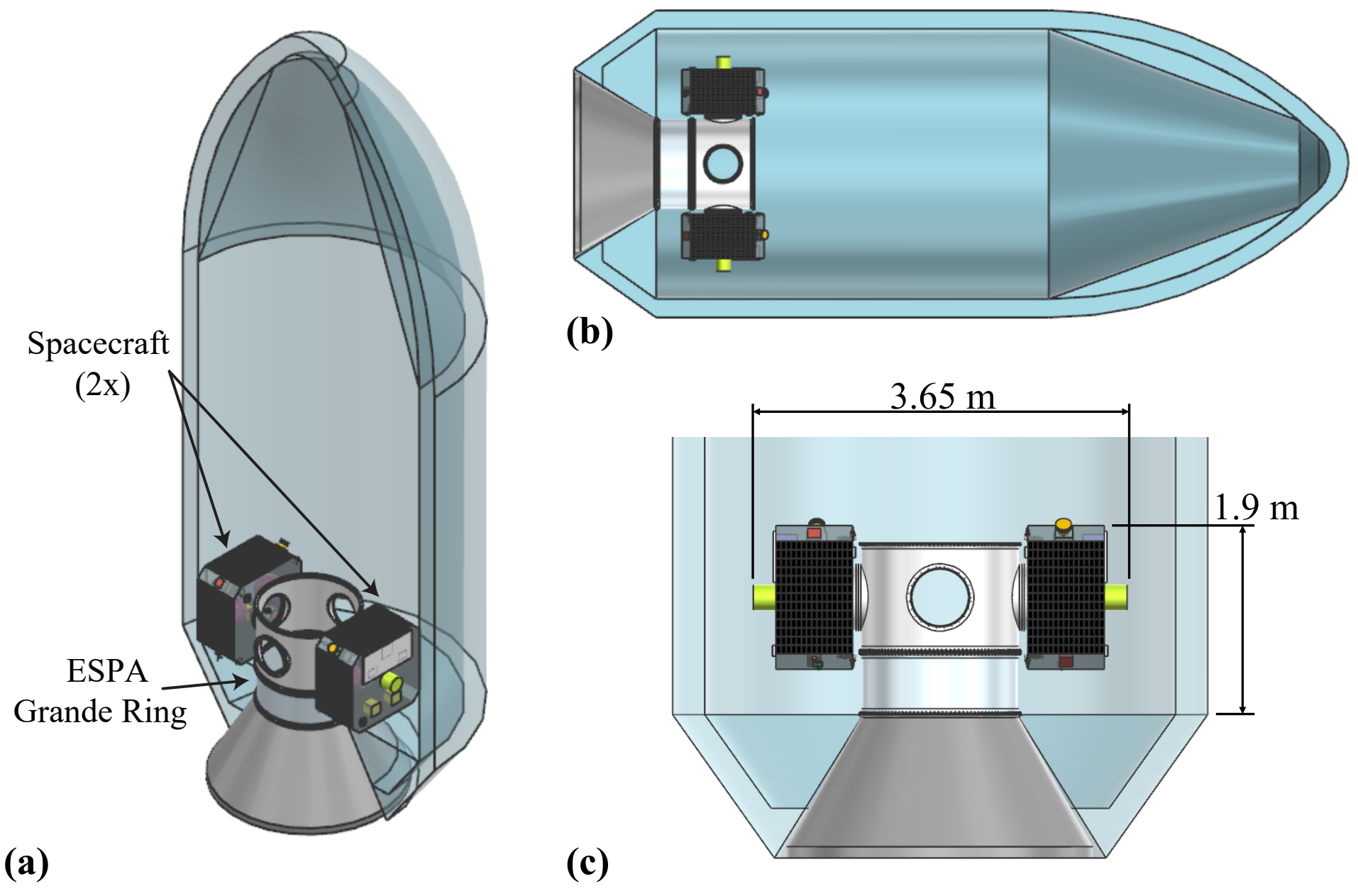

Figure 8: Three views of the spacecraft launch stack inside the 4.6 m launch vehicle fairing.

## 5. Spacecraft Design

### *5.1. Overview*

Renderings of the CRIMP spacecraft with key components labeled are shown in Figure 9. The two spacecraft are identical. The spacecraft bus is a custom design featuring body-mounted solar panels, a deployable magnetometer boom, and a scan platform for the REPTile instrument. The remaining components are internal to the spacecraft structure with the exception of the low gain antenna, the sun sensors, the HPCA, and the Ion ESAs.

The predicted dry mass for each spacecraft is 237.8 kg with an allowable dry mass of 282.1 kg. The latter is carrying 19% margin per the method specified in the 2019 AO (i.e., $margin = \frac{Allowable\ Dry\ Mass - Predicted\ Dry\ Mass}{Predicted\ Dry\ Mass}$). Note that the allowable dry mass is calculated via the sum of the predicted mass (i.e., the margin is carried on the entire spacecraft, not the individual subsystems—which each have a mass growth allowance). The propellant mass was sized to the allowable dry mass, resulting in 60.6 kg of propellant and a predicted wet mass of 298.4 kg. The allowable wet mass was found to be 342.7 kg. The allowable launch mass was determined to be 1202.9 kg via the summation of the allowable wet mass for two spacecraft, the ESPA Grande ring, and the separation system.

In addition to the scientific instruments, the essential spacecraft subsystems include: propulsion, attitude determination and control (ADCS), telecommunications, thermal control, command and data handling (C&DH), flight software, power, and the mechanical system (including spacecraft structure). Table 3 summarizes the predicted mass (i.e., after inclusion of any relevant mass growth allowance) and maximum operating power (i.e., in one of the six operating modes, see Sec. 5.7) after initial deployment for each of the spacecraft subsystems with the exception of the flight software. The predicted mass for the required cabling (i.e., wire harness) is also included.

The following subsections provide details, including key trades, about each of the spacecraft subsystems as well as the ground system and data archiving plan. Note that all individual component selections and design decisions were made in collaboration with NASA JPL's Team-X.

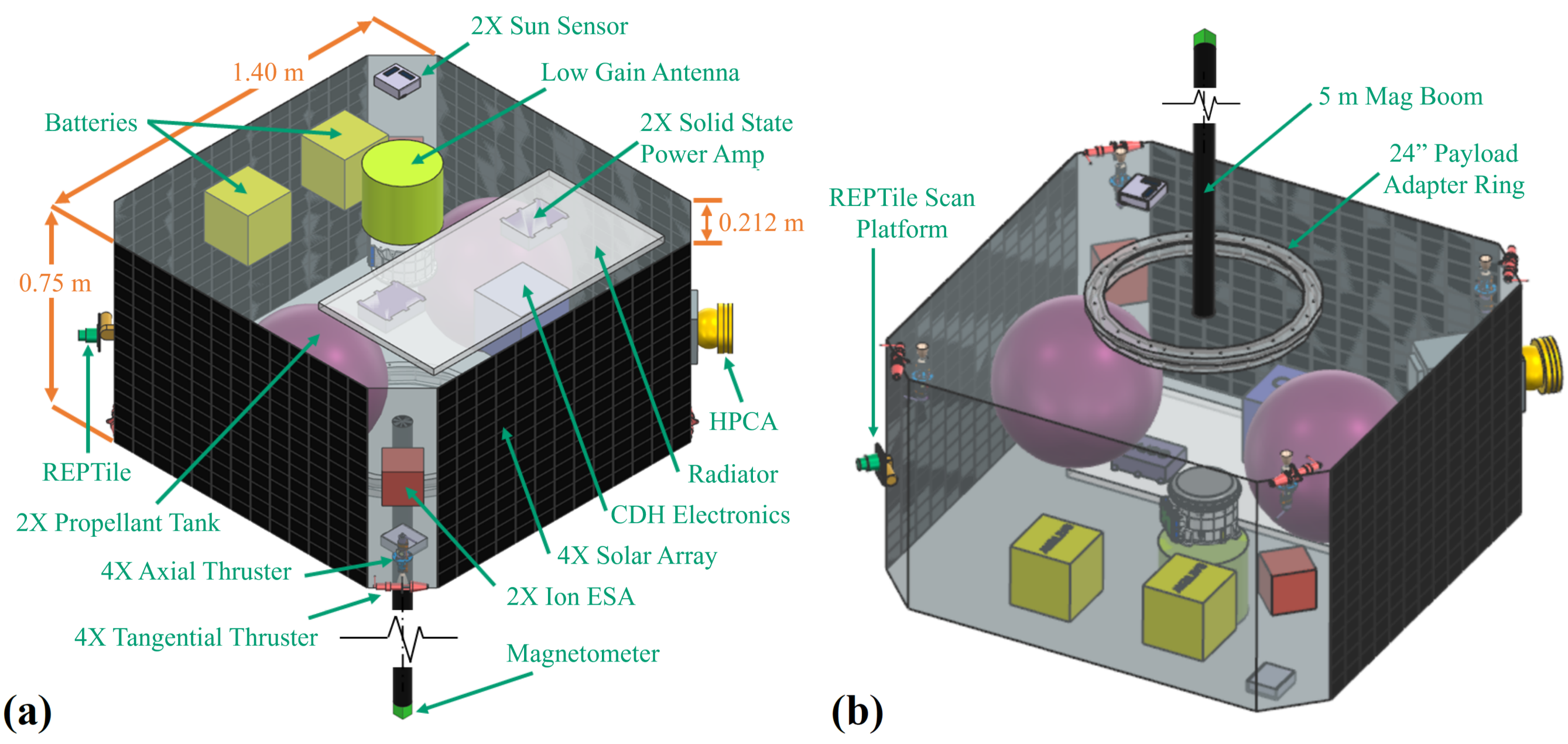

Figure 9: Spacecraft Configuration Schematic. Panel (a) shows the spacecraft with the top deck up. Panel (b) depicts the spacecraft with the bottom deck up.

Table 3: Predicted mass and maximum operating power for each subsystem

| Subsystem | Predicted Mass (kg) | Max. Power (W) |
|---|---|---|
| ADCS | 4.6 | 12.5 |
| C&DH | 8.1 | 22.6 |
| Power | 52.1 | 3.3 |
| Propulsion (Dry) | 21.8 | 78.7 |
| Mechanical | 73.6 | 0 |
| Telecommunications | 9.8 | 32 |
| Thermal | 25.6 | 40.7 |
| Instrumentation | 21.2 | 22.2 |
| Cabling | 19.2 | — |
| Total Dry Mass | 236.2 | — |
| Pressurant | 1.6 | — |
| Propellant | 60.6 | — |
| Total Wet Mass | 298.4 | — |

*5.2. Propulsion System*

A hydrazine monopropellant blowdown system was selected for orbital maneuvering, attitude adjustments, and spin-up and spin-down of the spacecraft. All components in the subsystem are TRL 9, with all hardware having previously been flown and proven through successful mission operations. A blowdown system was chosen for ease of operations and system simplicity. Duplicate pressure transducers and latch valves are implemented for redundancy, allowing the thrusters to be used in the event of a mechanical and/or sensor failure. Figure 10 provides a schematic of the proposed propulsion system.

The propulsion system was sized to an allowable wet mass of 343 kg and an allowable dry mass of 282 kg. This corresponds to approximately 60.6 kg of propellant with 0.1 kg of pressurant. All propellant will be held in two identical 80512-1 ATK PSI tanks with a total volume of 0.06 m$^3$ (3,660 in$^3$). A tank ullage of approximately 25% was assumed. The beginning-of-mission tank pressure will be 2,757.9 kPa (absolute, 400 psia). This decreases to 979 kPa (absolute, 142 psia) at the end of the mission. This end-of-mission pressure provides sufficient operational margin to the minimum allowed pressure needed by both types of thrusters discussed below. A residual propellant holdup of 2.5% is calculated for the end of the mission.

For orbital maneuvering and attitude adjustment, four MR-106L (5.0 lbf) thrusters with a specific impulse of 233 sec were selected. Over the length of the primary mission, these thrusters are anticipated to provide approximately 339 m/s of $\Delta V$. For spin-up and spin-down operations, each spacecraft also has four MR-103G (0.2 lbf) thrusters with a specific impulse of 223 sec. The thrusters are not used for attitude control because the spacecraft is spin-stabilized as discussed in the following subsection.

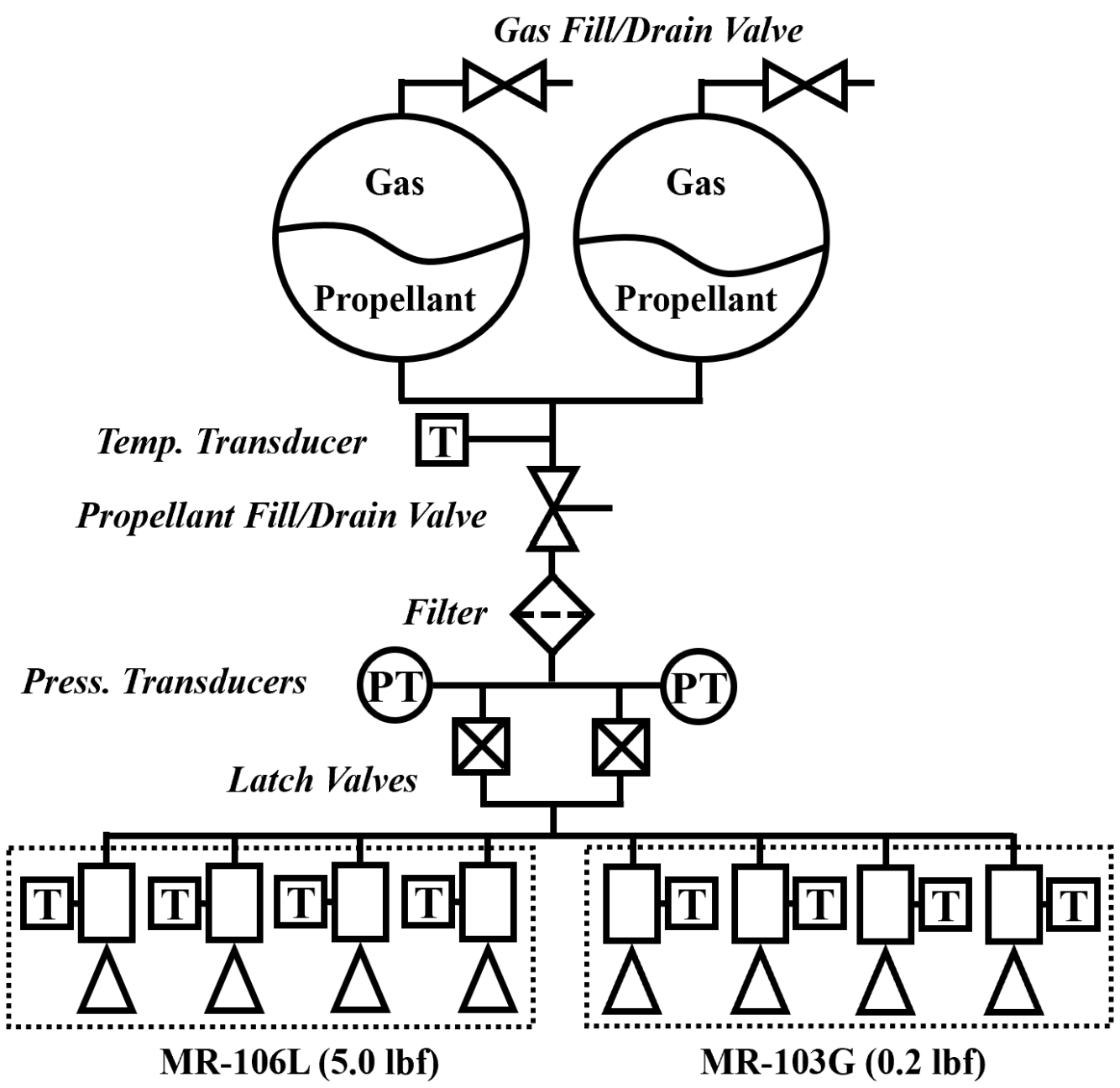

Figure 10: Propulsion System Schematic. The values in parenthesis are the individual thruster thrust values.

*5.3. Attitude Determination and Control System*

Science Objectives 1 and 2 drove the selection of a spin-stabilized approach to attitude control. To meet the time resolution requirements, both CRIMP spacecraft are required to rotate at a rate greater than or equal to 15 revolutions per minute around the vertical axis in Figure 9. Spin-stabilization also helps ensure that the spacecraft have a consistent orientation with respect to the nominal reconnecting magnetic field components in the magnetosphere. This is particularly important for the fluxgate magnetometer and plasma instruments, which require steady, predictable pointing to collect high-fidelity scientific data. Spin-stabilization has been widely used in heliophysics missions [87], particularly for spacecraft in geocentric orbits, as it minimizes disturbances, reduces control complexity, and improves long-term stability without requiring continuous attitude corrections.

The ADCS for CRIMP is tightly coupled with the propulsion system, as the initial spin-up and spin-down maneuvers rely on dedicated thrusters. The spacecraft employs four MR-103G thrusters for spin-up, spin-down, and spin adjustments. Once spin-stabilized, the spacecraft maintains its orientation using a combination of passive stability and active control through magnetic torque rods, which interact with Earth's geomagnetic field to make fine attitude adjustments as needed.

The CRIMP spacecraft will be equipped with a multi-sensor attitude determination system. This includes two sun sensors for coarse orientation reference, an inertial measurement unit (IMU) for real-time rate sensing, and the fluxgate magnetometer for determining orientation relative to Earth's magnetic field. The fluxgate magnetometer serves a dual purpose, contributing both to scientific data collection and as a means for determining the orientation of the spacecraft. This collection of sensors allows the spin rate of each spacecraft to be actively monitored. Adjustments to the spin rate can be made with a set of three magnetic torque rods or, if necessary, small propulsive maneuvers.

Several alternative ADCS architectures were considered during the mission design phase. A three-axis stabilized system was evaluated for its potential to provide potentially higher pointing accuracy; however, it was ultimately rejected due to increased complexity, mass, and power consumption. Additionally, a despun platform was explored to provide a stable, non-rotating reference frame for the communications subsystem, but this was deemed impractical given the cost, mechanical risk, and added complexity of maintaining said system. Spinning down for communications was also considered. This, however, was found to be overly propellant-intensive (as every communications pass would require spinning the spacecraft down and then up again before resuming science operations). Instead, the selection of a fully spin-stabilized configuration follows the practice used in previous missions such as THEMIS [88] and MMS [89]. This design decision balances simplicity, robustness, and efficiency while enabling the science objectives to be met. Note that additional filtering and system analysis is required to ensure the pointing knowledge and control requirements listed in Table 2 can be met.

The predicted mass of the ADCS (neglecting the propulsion components detailed in the previous section) is 4.6 kg. The three magnetic torquers are anticipated to have a total mass of 2.6 kg while the IMU has a predicted mass of 1.5 kg. The remaining 0.5 kg corresponds to the sun sensors.

### *5.4. Telecommunications and Ground System*

The telecommunications system consists of a UST-Lite (Universal Space Transponder) operating in the X-band for both receiving and transmitting signals. The system includes an external 1 W Solid-State Power Amplifier (SSPA) to amplify the signal before transmission, a diplexer to separate the uplink and downlink signals, and a low-gain antenna (LGA) for receiving and transmitting signals in the X-band (see Figure 11). The total mass of the subsystem is 9.8 kg, including a mass growth allowance, which accounts for potential increases in mass due to design modifications or hardware refinements. The UST-Lite transponder has a predicted mass of 3.3 kg while the LGA has a predicted mass of 2.1 kg. The diplexer and amplifier have predicted masses of 0.9 kg and 0.8 kg, respectively. The remaining mass comes from various minor components, connectors, and cabling. The proposed telecommunications system meets the downlink requirement with a >3 dB margin and the uplink requirement with a >10 dB margin. The system's power consumption is 32 W in transmit and receive mode, and 22 W in receive-only mode.

During the design process, other antenna options, such as a patch antenna, were also considered. However, patch antennas are typically directional and require precise pointing to maintain a stable connection with the ground station. To ensure continuous coverage with the spacecraft spinning, multiple patch antennas would have been necessary, increasing the mass, power consumption, and system complexity. Instead, an LGA mounted on the spin axis provides wide beam coverage. This eliminates the need for precise pointing and offers a simpler, more robust design for the spinning spacecraft.

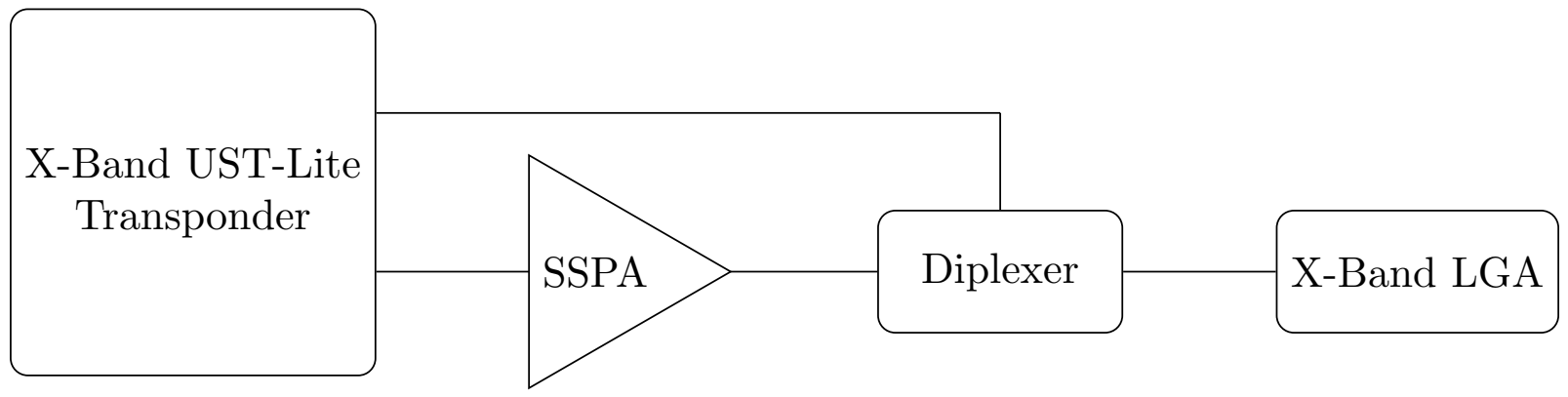

Figure 11: Telecommunications Subsystem Architecture

Data downlinking and uplinking will both use X-band communications. While the Ka-band would enable faster downlinking, it has a more narrow bandwidth window than X-band and thus would require a more accurate pointing capability from the spacecraft's antenna. The X-band permissible data downlink rate is 300 kbps (0.018 Gbits/min) and the science objectives require 0.34 Gbits of compressed data from the spacecraft to be downlinked per orbit. Therefore, approximately 20 minutes are required to transmit all data. To improve robustness to missed passes and other delays in data transmission, 30 minute passes are currently planned. The capacity for a 30 minute pass is calculated to be 0.54 Gbits. Thus, only 63% (0.34 Gbits/0.54 Gbits) of the total capacity is planned for use during standard operations. Data uplink is less demanding, with a required uplink rate of 2 kbps. Note that safe mode (see Sec. 5.7) can support lower data rates, as only spacecraft health information will be transmitted. Safe mode will downlink at 50 kbps at maximum range in the X-band, while the uplink rate will remain 2 kbps. Additionally, in order to mitigate a possible mission risk, a telecommunications downlink performance study is planned to better understand the throughput impacts of the spinning antenna (fixed to the spacecraft).

Due to the 22° orbital inclination, the ground data systems (GDS) facilities used will be mid- to low-latitude receivers in the Near Space Network (NSN). This includes, but is not limited to, Wallops Island (USA), which is located at 38° N and has an 11.3 m diameter antenna. Other options include White Sands (USA), located at 32° N, and Santiago (Chile), located at 34° S. With the inclusion of southern hemispheric receivers at places such as Santiago, there are abundant opportunities to communicate with the spacecraft throughout the standard mission profile.

The scope of operations for CRIMP is typical for a mission operations center. Nominally, there will be two mission support areas: one for operations at the NASA JPL Mission Support Area and one for assembly, test, launch, and operation (ATLO). The latter will be used for spacecraft assembly and testbeds for flight software development.

The flight software will be written for 10 day sequencing and will be updated weekly (i.e., the spacecraft will have 10 days of instructions on board that are updated every 7 days). The NASA Advanced Multi-Mission Operations System (AMMOS, see Ref. [90]) ground software is currently planned for mission operations. CRIMP carries 30% cost reserves in mission operation systems (MOS) and GDS in order to increase robustness and to be mission resilient in cost during both development and operations.

### *5.4.1. Data Processing and Archiving*

As previously noted, closure on the science objectives requires approximately 0.34 Gbits of compressed data to be downlinked from each spacecraft per orbit. This data will be generated over the nominal 6-11 $R_E$ interval or an equivalent interval selected by a scientist-in-the-loop (SITL) similar to the current MMS mission implementation (see Ref. [20]). This nominal data volume assumes a compression rate of 4x. Increasing the compression to 10x for CRIMP's plasma data may also be possible using methods similar to those developed for the MMS mission (e.g., Ref. [91]). This approach would allow for a greater volume of data to be selected and downlinked for each orbit, providing additional opportunities for a guest investigator program.

After downlinking, the raw science data (L0) will be processed through the AMMOS software into level 1-4 science data.[3] As part of this process, the data will be rotated and transformed into spacecraft and Geocentric Solar Ecliptic (GSE) coordinates. The resulting science data would then be archived using NASA's Space Physics Data Facility (SPDF). For a full two year mission duration, this would result in ∼138 Gbits of science data to be stored in the SPDF. An outline of the data archiving plan is shown in Figure 12.

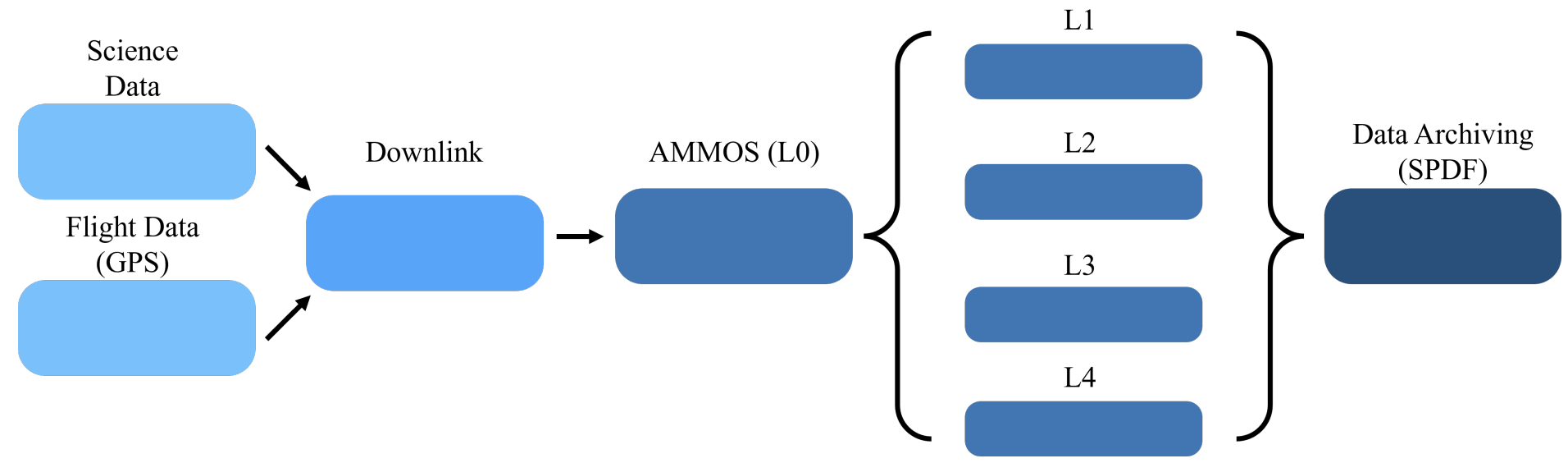

Figure 12: Outline of CRIMP's proposed data processing and archiving plan. Science data is downlinked at periapsis and then processed using the AMMOS software into level 1-4 science data before being stored using the SPDF.

### *5.5. Command & Data Handling System and Flight Software*

The Command and Data Handling (C&DH) system was designed using cost-effective, space-proven (TRL 9) components. The Sphinx processing unit was selected with this criteria in mind. This is a radiation-tolerant C&DH system based on the Aeroflex GR712 dual-core LEON3FT processor, capable of operating at 100 MHz with 2×134 DMIPS performance. The Sphinx integrates a Microsemi ProASIC FPGA, providing additional computational and control flexibility. Designed specifically for deep space environments, it follows a compact 10 cm × 10 cm form factor, has a mass less than 200 grams, and operates with 1.5 to 7 W of power. To ensure radiation resilience, its components are rated for at least 30 krad (Si) total dose, with the processor itself tolerant up to 300 krad (Si), and hardened against Single Event Latchups (SELs) with Linear Energy Transfer (LET) up to 37 MeV-cm$^2$/mg [92].

[3]PDS4 Processing Levels for Science Data Sets. See: `https://pds.nasa.gov/datastandards/documents/ProcessingLevelsMapping_220202b.pdf`, accessed 24 July 2025.

The C&DH system requires components for interfacing with the telecommunication system, managing incoming engineering telemetry (including magnetometer data), and power distribution. The system also includes 128 GB of storage, providing ample capacity to accommodate the estimated 42.5 MB of scientific data collected per orbit. Finally, the system architecture includes a backplane with six slots, providing modularity and potential expandability, as well as a chassis to house all components. The total predicted mass of the C&DH system is 8.1 kg. Note that the chassis has a predicted mass of 3.7 kg and is anticipated to be the most massive component in the C&DH system.

The selected Flight Software (FSW) has previously flown and will require only minor modifications for the CRIMP mission. The proposed FSW infrastructure is based on the F Prime[4] open-source modular, reusable, and adaptable framework developed at NASA JPL [93]. F Prime has flown in multiple missions and instruments such as the ISS-RapidScat instrument, the ASTERIA Cubesat, the Ingenuity Mars Helicopter, and the Lunar Flashlight and Near-Earth Asteroid Scout missions [94]. F Prime is composed of (1) a modular architecture that uses discrete components that communicate between them using ports; (2) a C++ framework that enables basic component functionalities such as message queues and threads; (3) tools for automatic generating code and defining the discrete components and their interconnections, with a large collection of already defined components available in the F Prime distribution; and (4) tools to test at unit and integration levels, including the F Prime Ground Support Equipment (GSE) software [93]. Additionally, the automatic code generator (known as the F Prime Autocoder) provides software tools for thread management, inter-process communication (IPC), commanding, telemetry, and parameters. This software architecture defines modular components with standard interfaces that allow rapid development, portability, reusability, analysis, and testing. This flexibility and adaptability support all the mission specific software requirements [94].

The F Prime Framework has minimal complexity and permits significant code reuse from previous missions. Specifically, the main engineering applications (i.e., telecommunication, thermal, and power subsystems) and the system services are on the current product line at JPL, requiring no substantial modifications. This is also the case for the GPS receiver data processing. Additionally, the payload accommodation (i.e., magnetometer, ion plasma suite, and electron telescope) will be simple to implement because the data handling is almost identical to the THEMIS and MMS missions. Only minor changes to the code from these missions would be required. On the other hand, the GN&C software will require major modifications because the particular spin-stabilized and propulsion methods planned for CRIMP are not on the current product line. Nevertheless, the spacecraft attitude control system has very low complexity and will require minimal development efforts. It is also relevant to note that the magnetometer data is used internally for GN&C purposes. Moreover, internal data management requires the implementation of background data compression processing while gathering instrument data and performing scientific data packetization and transmission.

[4] See: `https://fprime.jpl.nasa.gov/`, accessed 23 Jan. 2026.

*5.6. Thermal System*

A cold-biased method leveraging multi-layer insulation (MLI) was selected for thermal control. A block diagram of the thermal control system is shown in Figure 13. The radiator was sized to compensate for the worst-case-hot (WCH) scenario (i.e., when the spacecraft will be the hottest throughout its nominal mission). This puts the system in a cold-biased state for all other scenarios, requiring heaters to warm the hardware and maintain their functionality throughout the mission. The WCH scenario is anticipated to be when the spacecraft is at perigee on the dayside. Here, the spacecraft receives the highest thermal loads from the Earth (albedo and inferred radiation) and the Sun. Assuming a battery will be mounted to the radiator, the temperature of the radiator must be maintained between 273 K and 303 K. Keeping the radiator at 293 K requires an estimated radiator area of 0.5 $m^2$.

The worst-case-cold (WCC) scenario occurs when the spacecraft experiences an eclipse. Here heater power is needed for survival. During this scenario, the estimated heater power required is approximately 133.9 W. However, if louvers were implemented on the radiator, the heater power for the WCC scenario could be decreased. Using the current area of the radiator, two 20-blade louvers can be accommodated. With the louvers, the heater power needed in the WCC drops to approximately 22 W.

The predicted mass of the entire thermal system is 25.6 kg. This includes 9.3 kg of multi-layer insulation and a 6.8 kg radiator. The louvers have a predicted mass of 3.8 kg. The remaining mass comes from heat pipes (0.8 kg), thermostats (1.3 kg), temperature sensors (2.0 kg), heaters (1.0 kg), and other minor components.

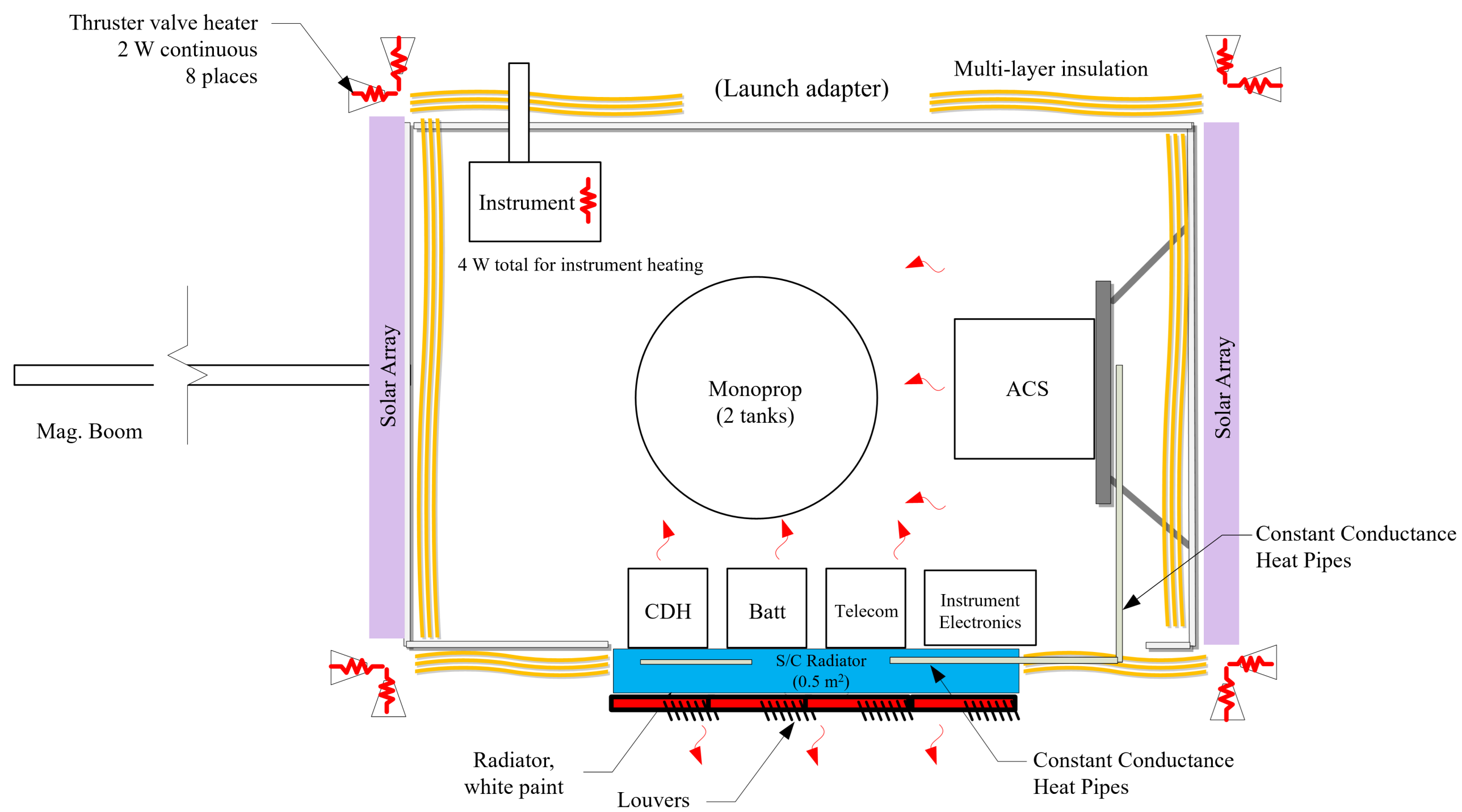

Figure 13: Block Diagram of the CRIMP mission's thermal system. The diagram depicts relevant systems and elements for temperature control. The diagram is not comprehensive, but conveys the key subsystem components.

*5.7. Power System*

Six power modes were identified as necessary throughout the nominal CRIMP mission: safe, trim maneuver, eclipse, science, downlink, and standby. Safe mode is triggered upon the spacecraft detecting a fault. In this mode, no science operations occur and the communications, attitude control, and telecommunications systems operate at maximum power while the thermal system operates at moderate power. Trim maneuver mode corresponds to the power required to conduct an orbital adjustment maneuver. In eclipse mode, no power is generated by the solar arrays and the heaters operate at maximum power. Science mode is the only power mode in which the scientific instruments are active. In downlink mode, data is transmitted from the spacecraft to the ground. Finally, standby mode accounts for standard house-keeping operations without collecting data (e.g., when the spacecraft is not on the dayside of the magnetopause). The current best estimate for the power required for each of these modes is summarized in Table 4. The rightmost column of Table 4 includes the current best estimate for the power required plus 43% contingency (i.e., $contingency = \frac{Allowable\ Power\ Required - Predicted\ Power\ Required}{Predicted\ Power\ Required}$). Note that while the trim maneuver mode requires the most power, this mode is only in use for short durations. Excluding safe mode, the remaining modes are more representative of standard operations.

The aforementioned power modes drove the selection of the power system components. The selected power system consists of a 2.8 $m^2$ Triple Junction GaAS solar array, two 1094 Wh Li-Ion batteries, and control electronics to provide the bus 32.8 V. The solar panels are body-mounted, non-deployable, and have a predicted mass of 12 kg. The solar array was assumed to produce approximately 234 W at beginning of life and 211 W at end of life. The solar arrays were sized for safe mode. Conversely, the batteries were sized for eclipse mode and assumed to have a 32% worst-case minimum state of charge. The combined predicted mass for the batteries was found to be 25.5 kg. The predicted mass for the entire power system (excluding cabling) was 52.3 kg.

Table 4: Power Modes

| Mode | Power Required (W) | Power Required + 43% Contingency (W) |
|---|---|---|
| Safe | 154.2 | 220.5 |
| Trim Maneuver | 173.8 | 248.5 |
| Eclipse | 100.9 | 144.2 |
| Science | 119.5 | 170.9 |
| Downlink | 90.6 | 129.6 |
| Standby | 100.9 | 144.2 |

*5.8. Mechanical System and Spacecraft Configuration*

The primary structure of the spacecraft bus will be custom designed to satisfy mission requirements for the instrument suite, spin rate, and launch vehicle fairing clearance. The CRIMP bus shares design elements with the THEMIS spacecraft, but features slightly larger dimensions to accommodate large solar panels. The spacecraft bus comprises a top deck, bottom deck, and eight vertical sides that form an approximate square with chamfered corners. The frame will be constructed from aluminum of various thicknesses greater than 2.5 mm.

Table 5 provides the predicted mass of each major element of the spacecraft structure and associated mechanisms, including the wiring harness. The top deck houses the low-gain antenna, radiator, and avionics, while the two batteries and two fuel tanks are located inside the bus. The 5 m carbon fiber deployable boom for the fluxgate magnetometer is mounted at the bottom of the bus. The expandable mesh boom is deployed from its stowed position during the spacecraft commissioning process.

The four solar panels are body-mounted on the major side faces of the bus, each with a surface area of 0.825 $m^2$. The instrument suite and tangential MR-103G thrusters are located on the corner panels. In addition, the REPTile instrument is mounted on a scan platform with a standard single rotating actuator. The scan platform will allow REPTile to have a pointing range of 44° - 133° from the spin axis and rotate the instrument ∼3.8° per month.

Table 5: Mechanical System Mass Breakdown

| Item | Predicted Mass (kg) |
|---|---|
| Primary/Secondary/Tertiary Structure | 38.2 |
| Power Support Structure | 2.5 |
| Telecommunications Support Structure | 0.3 |
| Balance/Ballast | 12.7 |
| Integration Hardware | 2.7 |
| Scan Platform Structure | 3.9 |
| Scan Platform Actuator Mechanism | 3.3 |
| Scan Platform Launch Locks Mechanism | 0.9 |
| Magnetometer Boom Mechanism (5 m) | 6.5 |
| Al Radiation Shielding (2.5 mm thickness, 1.363 $m^2$ area) | 2.6 |
| Mechanical Total | 73.6 |
| Harness Total | 19.2 |

*5.8.1. Radiation Shielding*

The CRIMP mission operates in a highly dynamic radiation environment, with an orbit extending from 1.3 $R_E$ at perigee to 15 $R_E$ at apogee. A significant portion of the radiation exposure occurs during the traversal of the Van Allen Belts, where the spacecraft encounters intense trapped electron radiation. These belts, composed primarily of energetic electrons and protons, pose a considerable risk to spacecraft electronics and require robust shielding strategies to ensure system reliability.

To assess the Total Ionizing Dose (TID) over the two-year mission duration, SPENVIS[5] was used for radiation modeling. The trapped radiation environment was evaluated using the AP-8 and AE-8 models (solar maximum), while the SAPPHIRE model, with a stormy magnetosphere, was applied for solar particle effects. The SHIELDOSE-2 model, assuming a spherical geometry, was used to compute radiation exposure for different shielding thicknesses.

[5]See: `https://www.spenvis.oma.be/`, accessed 23 Jan. 2026.

The results indicate that the TID varies from 4.5 krad (Si) with 10 mm of aluminum shielding to 31 krad (Si) with 5 mm of shielding. For a 6 mm aluminum shielding configuration, the estimated dose is approximately 15.7 krad (Si).

The primary radiation mitigation strategy for CRIMP will be spot shielding, where sensitive components receive additional localized shielding to optimize radiation protection while minimizing overall spacecraft mass. A total of 1.363 $m^2$ of 2.5 mm aluminum shielding was estimated to be required (i.e., in addition to the primary structure thickness). This results in a predicted shielding mass of 2.6 kg and an estimated total shielding thickness (i.e., primary structure plus additional shielding) of $\approx$5-6 mm. This approach ensures the longevity and robustness of the mission's critical electronics while maintaining efficient mass allocation. As the concept matures, additional detailed shielding analysis will be conducted to ensure that the spacecraft bus and its instrumentation can tolerate the expected radiation environment.

## 6. Cost Analysis

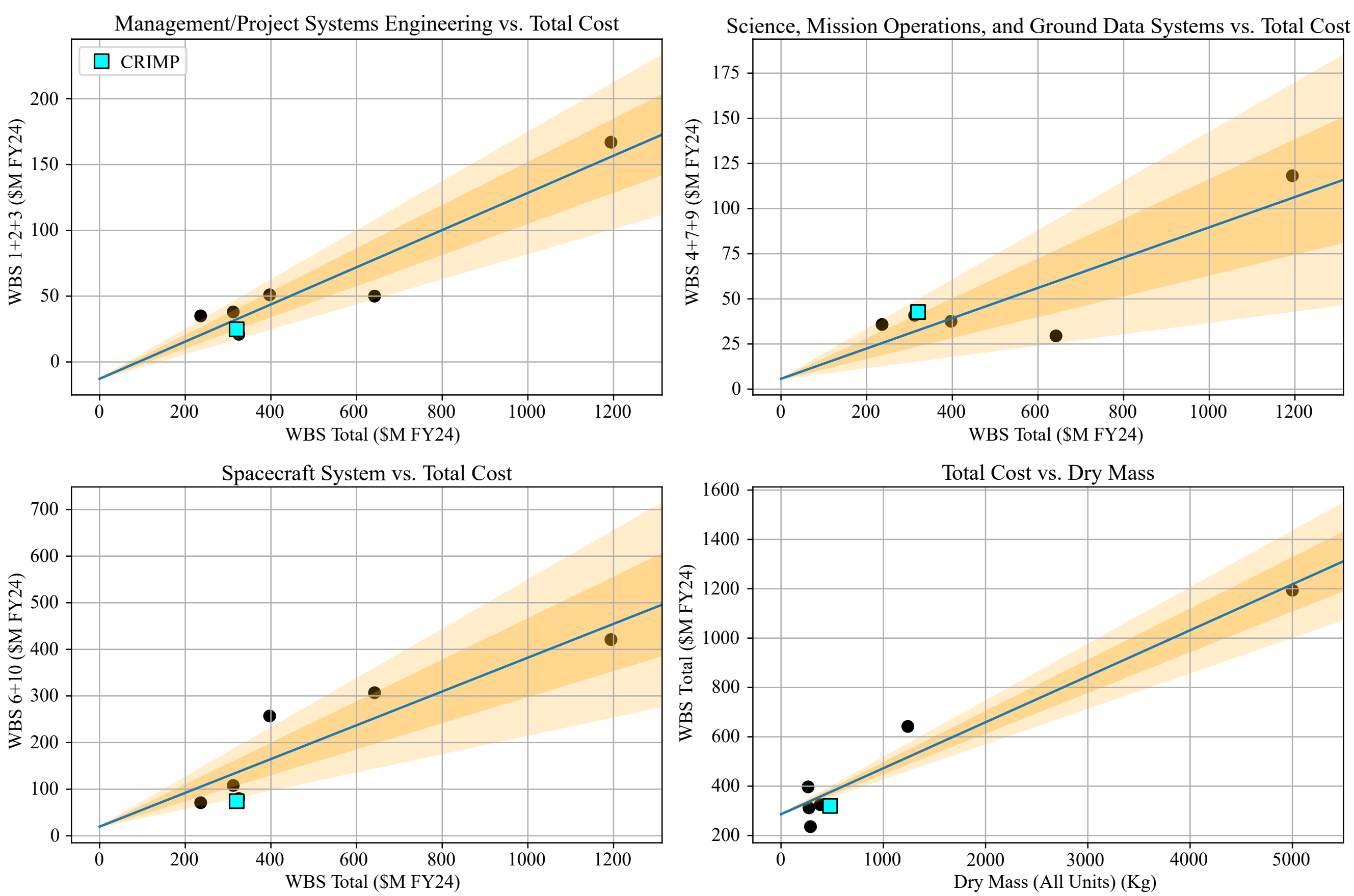

Figure 14: Validation of CRIMP Cost Estimates against historical data. For various comparisons of different WBS elements and mission parameters, the CRIMP CBE cost estimates (cyan square) are placed in context of historical trends. Other mission data (black circles) are drawn from prior NASA Astronomy and Heliophysics MIDEX missions, and higher class heliophysics multi-spacecraft missions. For each case, a linear regression line is shown along with $1\sigma$ and $2\sigma$ shading. CRIMP cost estimates are in family with historical trends.

The CRIMP mission was evaluated to be cost-feasible under a cost cap of $301M in FY24 USD with a 30% reserve posture, consistent with the requirements of the NASA 2019 Heliophysics MIDEX Announcement of Opportunity

Table 6: Work Breakdown Structure (WBS) Level 2 Current Best Estimate

| Work Breakdown Structure Element | Full Life Cycle Cost, $M FY24 (Phase A-F) | Basis of Estimate/Note |
|---|---|---|
| 1 Project Management | 6.6 | JPL ICM + Historical Validation |
| 2 Project Systems Engineering | 9.1 | JPL ICM + Historical Validation |
| 3 Safety and Mission Assurance | 10.1 | JPL ICM + Historical Validation |
| 4 Science | 11.2 | JPL ICM + Industry provided information |
| 5 Payload System | 42.5 | NICM + JPL ICM for management wraps |
| 6 + 10 Spacecraft Systems + ATLO | 74.2 | Estimated from historical mass to cost ratio data |
| 7 Mission Operations System | 13.7 | JPL ICM + Industry provided information |
| 8 Launch System | 0.0 | NASA Provided per AO |
| 9 Ground System | 20.6 | JPL ICM + Industry provided information |
| 12 Mission Design and Navigation | 3.3 | JPL ICM |
| Reserves | 78.4 | 30% Posture |
| Phase A | 1.5 | AO Requirement |
| PIMMC | 297.3 | |
| PIMMC + Contribution | 319.6 | 2nd identical unit provided by external partner |

with the important caveat that an identical (recurring-cost only) spacecraft unit is contributed (provided) by an external partner. It is anticipated that this requirement may be avoided with the further maturation of the design and cost analysis, falling in line with previous multi-spacecraft MIDEX cost cap missions.

Cost modeling was performed in collaboration with NASA JPL's Team-X using the NASA JPL Institutional Cost Model (ICM) and the NASA Instrument Cost Model (NICM, see Ref. [95]). Validation was performed with historical MIDEX-scale data from all NASA science mission directorate divisions, and from Heliophysics division missions of other classes. In all cases, the CRIMP cost was observed to be in family with historical trends (see Figure 14).

Instrument cost estimates were derived (via NICM) from analogous instruments in flight on board NASA's MMS mission, THEMIS mission, and the Van Allen Probes and with associated management and engineering costs derived from the NASA JPL ICM.

Including reserves and excluding the assumed external contribution, the current best estimated (CBE) PIMMC totals $297.3M (FY 24). A full breakdown by NASA Work Breakdown Structure (WBS, Level 2; see Ref. [96, Appendix C]) is given in Table 6. For this AO, WBS 11 (Education and Public Outreach) was not required and is thus omitted from Table 6.

Note that the cost information contained in this paper is of a budgetary and planning nature and is intended for informational purposes only. It does not constitute a commitment on the part of NASA JPL and/or California Institute of Technology.

## 7. Discussion

### *7.1. Links to the HSO*

The Heliophysics System Observatory (HSO) represents NASA's fleet of heliophysics missions and observational assets that allow for a combined view of the space weather system. These missions include upstream solar wind monitors such as the Advanced Composition Explorer (ACE; see Ref. [97]) and Wind [98–100] as well as magnetospheric missions such as THEMIS, MMS, and the recently launched Tandem Reconnection and Cusp Electrodynamics Reconnaissance Satellites (TRACERS; see Refs. [101, 102]) mission. Additionally, potential future NASA concepts, such as the the Links Between Regions and Scales in Geospace (Links) mission (a flagship-class systems science mission proposed by the 2024 Heliophysics Decadal Survey), would allow for spatial (0.5-1 $R_E$) and temporal investigations of the near-Earth environment, focusing on the coupling between the solar wind, magnetosphere, and ionosphere systems with a large-scale fleet of observing spacecraft.

Depending on the potential launch window for future Heliophysics MIDEX Explorer class missions, the current or future HSO would provide rich opportunities for collaboration. For example, utilizing MMS and CRIMP's ion mass spectrometer instrument along the magnetopause would allow for large scale explorations into magnetospheric ion outflows across the dayside and flank magnetosphere. Additionally, CRIMP would allow a mesoscale complement to MMS's kinetic scale observations in the magnetosphere, potentially allowing for direct comparisons between kinetic observations of magnetic reconnection and the mesoscale dynamics occurring at the magnetopause at that time. Recently launched missions, such as TRACERS, or future mission projects, like Links, would then allow for in depth analyses into plasma movement into and out of the magnetosphere system, widely covering the local, mesoscale, and global scales.

### *7.2. Guest Investigator Opportunities and Magnetotail Mesoscale Dynamics*

While CRIMP's motivating science objectives focus on the dayside phenomenological structures and plasma outflow along the magnetopause, there are a plethora of additional mesoscale science and opportunities in the magnetotail, bow shock, and magnetosheath for guest investigator opportunities. At the bow shock and magnetosheath, CRIMP could provide direct observations of HFAs and HSJs. While in the magnetotail, CRIMP could investigate mesoscale phenomena such as dipolarization fronts and dipololarization flux bundles as well as help study the localization of magnetic reconnection sites in the tail current sheet (e.g., Ref. [103]). Over the course of a two year mission, CRIMP would spend 453 days in the flank magnetosphere and tail regions where CRIMP could offer additional highly valuable mesoscale contemporaneous measurements and potential conjunctions with other HSO missions (THEMIS, MMS, etc.), creating the potential for enhanced community participation under a robust guest investigator program.

### *7.3. Future Development Directions*

This paper details a single iteration of the CRIMP mission concept that can complete compelling science within the constraints of the Heliophysics Explorers Program 2019 MIDEX AO. There are, however, several directions for future development, investigation, and study. The first of which is the inclusion of additional spacecraft (e.g., four instead of two). While the base number of spacecraft necessary to achieve the science objectives detailed here is two, additional spacecraft would increase the mission's redundancy, resiliency, and data robustness. Further, additional spacecraft may permit measurements to be taken in the $Z_{GSE}$ axis, allowing for investigation into the 3D nature of mesoscale phenomena along the magnetopause. These additional spacecraft would also allow for the exploration of different ranges of mesoscale separations, including expanded separations out to 5 $R_E$. This would permit the observation of larger structures on the magnetosphere's flanks such as KHI. Additional analysis and maturation of the design is required to determine if including more spacecraft is feasible (n.b., it may also be possible to eliminate the need for a contributed spacecraft).

The exploration of instrumentation specifically developed for CRIMP is another possibility for future work. While CRIMP's science goals can be achieved with the instrumentation detailed in Sec. 3, the inclusion of purpose-built instrumentation would allow for potential refinements in the mission's cost, allow for enhanced scientific discovery, and increased data robustness. This includes the potential to expand CRIMP's science goals to target additional magnetospheric outflows, such as cold plasmas, or to directly target the detection of magnetic reconnection sites.

Finally, additional development may enable exploration of mesocale dynamics in the magnetotail. In this current iteration, CRIMP focuses exclusively on dayside mesoscale dynamics and particle outflows; however, there are a myriad of additional mesoscale phenomena that occur in the magnetotail and in the nightside magnetosphere. While CRIMP currently plans to offer a guest observer program to enable data collection outside the primary scientific region of interest, CRIMP could aim to address expanded science objectives 1 and 3 that explore magnetospheric plasma populations, their movements, and mesoscale structures in the nightside magnetosphere (e.g., dipolarization fronts, the localization of X-lines in the magnetotail current sheet, etc.; see Ref. [103] and the references therein).

## 8. Conclusion

The process of energy transfer in the near-Earth environment is a critical aspect of understanding space weather events and their impacts on modern infrastructure. The Earth's magnetopause holds a key role in this process, acting as the entry gate of the solar wind's energy into the magnetosphere through the process of magnetic reconnection. While the global and kinetic scale aspects of the magnetopause and magnetic reconnection have been widely studied, the impacts of mesoscale phenomenological structures, drivers, and plasma outflows along the magnetopause are fundamental open questions that require innovative, targeted investigations to resolve.

As modern society's reliance on space-based assets grows, the ability to accurately determine this energy inflow into the near-Earth environment is crucial. Not only is the space-based economy expanding and projected to encompass over $1.8 trillion by 2035 [104], but humanity is also striving to both reestablish a presence on the Moon and potentially reach new frontiers on Mars. In order to protect these assets and to understand the impacts of space weather systems at other (induced or intrinsic) magnetospheres, we must understand the impacts of these systems and how they interconnect across scales as highlighted by the 2024 Heliophysics Decadal Survey's PSG-1 and SG-1a. Mission concepts that target these science goals and the mesoscale dynamics connecting global drivers with local energetic processes will fundamentally help to protect these efforts by clarifying our understanding of the emergent space weather system. The Compression and Reconnection Investigations of the Magnetopause (CRIMP) mission concept is designed to fill this role through a spacecraft configuration that allows multipoint, contemporaneous measurements at the Earth's magnetopause. Through these measurements, CRIMP will determine the spatial scale size, extent, and temporal evolution of energy and mass transfer processes at the magnetopause—enabling novel

understanding of how the solar wind energy input to the magnetosphere is transmitted between regions and across scales.

CRIMP targets three science objectives that address the aforementioned science goals:

- **Objective 1**: Determine if the global dayside reconnection rate is reduced by localized enhancements in mass density of magnetospheric plasma ($H^+$, $He^+$, $He^{++}$, $O^+$) at the magnetopause boundary.
- **Objective 2**: Determine if the magnetopause boundary structure is driven by global-scale solar wind interactions or by local-scale conditions in the magnetosheath.
- **Objective 3**: Determine if the loss of ultra-relativistic radiation belt electrons across the dayside magnetopause is due to the gradient drift of electrons across the magnetic discontinuity or magnetopause dynamics.

This study is a point design existence proof showing that CRIMP is a technically- and cost-feasible, hypothesis-driven mission concept capable of addressing these science objectives within the constraints of the Heliophysics Explorers Program 2019 MIDEX AO. The use of twin spacecraft in precisely phased orbital lobes allows multipoint, contemporaneous measurements at the magnetopause spaced 1-3 $R_E$ apart. Each spacecraft will be equipped with an identical suite of magnetometers, ion mass density and ion spectrometers, and high energy electron telescopes that allow CRIMP to resolve magnetospheric outflows, magnetopause mesoscale structures, and radiation belt losses along the dayside magnetosphere. This orbital configuration and scientific suite enables unprecedented mesoscale observations along the dayside magnetopause and has the potential to revolutionize our understanding of energy transfer in the magnetosphere.

## Appendix A. Mission Formulation and Design Contributions

This mission concept and manuscript represent the combined efforts of a highly motivated team of 18 early-career researchers. All participants made valued contributions to the final product. Significant work was completed prior to the culminating week of the 2024 NASA Heliophysics Mission Design School school involving subgroup projects, late nights, and many collaborative problem solving sessions. Participants worked in subgroups such as science, engineering, publication, instrumentation, and data management. Further, several team members held "Hex" roles which were intended to mirror traditional mission formulation leadership roles. Near the end of the school and during the culminating week with NASA JPL's Team-X, each participant was also responsible for a specific mission development role or subsystem. Hex roles, culminating week system specializations, and other areas of major contributions are listed for each participant below (n.b., the participants are listed alphabetically). Further analysis and refinement of the mission concept, design, and analysis was continued after the completion of the formal mission design school.

**Samuel Badman**: Cost. **Jason Beedle**: Principal Investigator and Objective 2 Lead. **Humberto Caldelas**: Propulsion System and Mission Design, Analysis, and Navigation. **Kelly Cantwell**: Proposal Manager and Deputy Systems Engineer. **Bryan Cline**: Lead Systems Engineer and Mission Design, Analysis, and Navigation. **Alex Hoffmann**: Power System. **Christian Hofmann**: Telecommunications and Mission Design, Analysis, and Navigation. **India Jackson**: Attitude Determination and Control System. **Tre'Shunda James**: Thermal System. **Miguel Martínez-Ledesma**: Flight Software. **Bruno Mattos**: Command and Data System and Radiation Analysis. **Brett McCuen**: Science Chair and Deputy Principal Investigator. **Sophie Phillips**: Ground Systems. **Bryan Reynolds**: Instruments. **Julie Rolla**: Capture Strategist. **Orlando Romeo**: Mechanical/Configuration. **Frances Staples**: Project Manager and Objective 3 Lead. **Michael Starkey**: Data Sufficiency Lead, Objective 1 Lead, and Mission Design, Analysis, and Navigation.

### Acknowledgments

This research was carried out at the NASA Jet Propulsion Laboratory/California Institute of Technology, and was performed under the 2024 Heliophysics Mission Design School (HMDS) and sponsored by the National Aeronautics and Space Administration (80NM0018D0004). The cost information contained in this paper is of a budgetary and planning nature and is intended for informational purposes only. It does not constitute a commitment on the part of NASA JPL and/or California Institute of Technology.

The authors sincerely thank Leslie Lowes, Joyce Armijo, Kevin Frank, Troy Hudson, and all of NASA JPL's Team-X for their support throughout the 2024 Heliophysics Mission Design School. We also thank the reviewers of our culminating HMDS presentation for their time and feedback. Finally, Bryan Cline thanks Zoë Woodworth for her help with Figure 6.